\documentclass[review]{elsarticle}

\usepackage{lineno,hyperref}
\usepackage{subfig}
\usepackage{graphicx}
\usepackage{color}  
\usepackage{dcolumn}
\usepackage{amsmath}
\usepackage{amsfonts}
\usepackage{fullpage}
\usepackage{bm}
\usepackage{todonotes} 
\usepackage{algorithm}
\usepackage{algorithmicx,algpseudocode}
\usepackage{soul} 

\usepackage{lipsum}
\makeatletter
 \def\ps@pprintTitle{%
 \let\@oddhead\@empty
 \let\@evenhead\@empty
 \def\@oddfoot{}%
 \let\@evenfoot\@oddfoot}
\makeatother

\begin{document}

\graphicspath{{Figures/}}

\begin{frontmatter}

	\title{A new approach for electronic heat conduction in molecular dynamics simulations}
	\author{Mohammad W. Ullah   and }
	\author{Mauricio Ponga$^*$}
	\address{Department of Mechanical Engineering, University of British Columbia, 2054 - 6250 Applied Science Lane, Vancouver, BC, V6T 1Z4, Canada}
	\cortext[mycorrespondingauthor]{Corresponding author}
	\ead{mponga@mech.ubc.ca}

	\begin{abstract}
		We present a new approach for the two-temperature molecular dynamics (MD) model for coupled simulations of electronic and phonon heat conduction in nanoscale systems. The proposed method uses a master equation to perform heat conduction of the electronic temperature eschewing the need to use a basis set to evaluate operators. This characteristic allows us to seamlessly couple the electronic heat conduction model with molecular dynamics codes without the need to introduce an auxiliary mesh. We implemented the methodology in the Large-scale Atomic/Molecular Massively Parallel Simulator (LAMMPS) code and through multiple examples, we validated the methodology. We then study the effect of electron-phonon interaction in high energy irradiation simulations and the effect of laser pulse on metallic materials. We show that the model provides an atomic level description in complex geometries of energy transfer between phonons and electrons. Thus, the proposed approach provides an alternative way to the two-temperature molecular dynamics models. The parallel performance and some aspects of the implementation are presented.
	\end{abstract}


\end{frontmatter}


\section{Introduction}

Classical molecular dynamics (MD) simulations have been used widely in materials physics to study the lattice and defect dynamics at the atomic scale in both equilibrium and non-equilibrium conditions. In classical MD, the thermal conductivity due to lattice contribution is considered, since the phonons are explicitly resolved. This makes MD simulations one of the most used methods to measure thermal properties of materials, where heat is mainly transported by phonons. Unfortunately, the electronic contribution is neglected in MD, and thus, the study of materials where electronic heat transport is important cannot be performed with accuracy in MD. This render MD simulations well suited but also essentially limited to the simulation of low level of electronic excitation and ionization events, semiconductors and insulators. For example, phenomenon like swift heavy ion irradiation, short-pulse laser interaction, and metal-semiconductor joints where electronic energy loss/gain is significant cannot be accurately described using MD.

A preferred method to address this issue is the two-temperature molecular dynamics (2T-MD) model. In this approach, a Fourier law is used to describe  the electronic energy transport and is combined with classical MD in a concurrent multiscale scheme. In this model, the temperature evolution in electronic subsystem is described by a heat diffusion equation \cite{kag:1957,Caro89,Fin91,Hak93,iva:2003},

\begin{equation} \label{eq:TTM_electron}
	C_e (T_e)\frac{\partial T_e}{\partial t} = \nabla[K_e(T_e)\nabla T_e] - G(T_e-T_l),
\end{equation}
where $C_e$, $K_e$, $T_e$, and $T_l$ are electronic heat capacity, thermal conductivity, local electronic and ionic temperature, respectively. $G$ is a electron-phonon coupling constant between electron and phonons.

One appealing aspect of the 2T-MD model is that it can concurrently simulate energy exchange between phonons and electrons during a simulation. The implementation of Eq. \ref{eq:TTM_electron} in MD codes has been achieved by a number of means, including finite differences (FD) \cite{iva:2003,duf:2007,sch:2002,Rutherford:2007} and the finite elements method (FEM) \cite{Wagner:2008}. In implementations using FD, the atomic system is divided into small cells (\emph{voxels}) that generate a mesh and the FD method is used to evaluate the Laplacian of the electronic temperature and solve Eq. \ref{eq:TTM_electron} for each voxel. Similarly, in implementations using FEM, the MD system is embedded into a FEM mesh, where some of the atoms generate the nodes of this mesh. Thus, the coupling is done by passing the MD simulation data into the FEM mesh, which solves Eq. \ref{eq:TTM_electron}. Other methodologies have coupled both phonons and electrons using a heat conduction equation as Eq. \ref{eq:TTM_electron} while completely removing the lattice dynamics, leading to free-energy type methods to describe thermo-mechanical response of materials \cite{Ariza:2012,Ponga:2012,Ponga:2015,Ponga:2017}. The 2T-MD approach has been successfully applied to multiple problem of technological interest such as, strong electron-phonon nonequilibrium problems \cite{Lin:2008}, laser ablation \cite{Leveugle:2004,Zhigilei:2009,shu:2017}, shock induced melting \cite{PhysRevB.74.012101}, metals and semi-metals interfaces \cite{Jones:2010},  energy transport across semiconductor-metal interfaces \cite{PhysRevB.85.205311}, and to irradiation damage of materials \cite{Zarkadoula:2016,Zarkadoula:2017,lno:2018,Sellami:2019,Jin:2019,Zarkadoula:2019}, to mention but a few.


Regardless the choice of the method to evaluate the Fourier law for the electronic temperature, previous implementations have been using finite size sub-divisions of the system and linked the behavior of a cluster of atoms into the FD grid or FEM nodes. However, in many situations of interest, large temperature gradients appear in short length and time scales, and very fine spatial discretization of the electronic mesh is required to properly describe the temperature variation. This indicates that heavily excited atoms and electrons can quickly exchange energy in \emph{local} regions of the system and thus, small resolution is required to resolve this issue. Hence, recent studies have been indicated the need for local 2T-MD methods that can accurately account for this \cite{Zarkadoula:2016,Zarkadoula:2017}. Additionally, in system where liquid/solid interfaces appear, the amorphous nature of the atoms make it difficult to be described with accuracy using a voxel description or a FEM mesh.


In this work, we develop a new version of the two-temperature molecular dynamics method by replacing the classical Fourier law by a Fokker-Planck equation to simulate the exchange of energy between electrons. The Fokker-Planck equation, used in the form of a \emph{master equation}, computes the probability that two adjacent electronic sites exchange energy through electronic heat flow. This eschews the need of using an auxiliary mesh in order to evaluate the Laplace operator and retains, at the same time, the smallest resolution that one can achieve in MD systems, e.g., the atomic resolution. This is important as our approach can reproduce complex geometries in comparison with FD or FEM meshes. Another remarkable aspect of this approach is that by virtue of the master equation, the electrons exchange energy in a \emph{local} fashion, which allows the application of our method to situations where the system is far from the thermodynamic equilibrium, e.g., large temperature gradients in small regions. Additionally, since the method does not need an auxiliary mesh, it can be seamlessly coupled to MD codes and the electronic temperature can be easily initialized by assigning properties to the atoms, which reduces the effort of the users to define the coupled electrons-phonons system. Hence, we implemented the new method into the LAMMPS code \cite{Lammps} and refer to it as ${\ell}$2T-MD, and make it available through its online site \cite{l2T-MD:website}.

The manuscript is organized as follows. First, we start by introducing the formulation and performing the energy balance to determine the coupling between electrons and phonons. Next, we demonstrate the methodology by using a simple one-dimensional example using the master equation approach to simulate the electronic and phonons heat conduction. 
We investigate the conservation of the energy in the system and the energy exchange between electrons and phonons. The parallel performance of the method is also studied. Finally, we study two problems of technological interest, an irradiation damage problem in Ni using a 50 KeV cascade simulations; and a laser spallation simulation in Ni. We conclude the manuscript with the main outcomes of this work.

\section{Methodology} \label{Section:Methodology}
We now describe the ${\ell}$2T-MD method. Let us now consider a system of $N$ particles at finite temperature. This temperature is the contribution of two players, e.g., phonons \emph{and} electrons. MD simulations explicitly model the phonons as a result of atomic interactions accounted for by the interatomic potential. Unfortunately, the electronic contribution to the temperature is neglected and difficult to model. The two-temperature model assumes that the temperature field can be split into two interacting subsystems, e.g., phonons and electrons. Thus, one has an electronic temperature, $T^e$, and a lattice temperature given by the \emph{classical} definition, i.e., $T^{\text{lat}} = \frac{K}{\frac{3}{2} k_B N}$, where \textcolor{black}{$K = \displaystyle \sum_{i=1}^N \frac{1}{2} m_i {\bf v}_i\cdot {\bf v}_i$} is the total kinetic energy of the atoms, $m_i$  and ${\bf v}_i$ are the  mass and the velocity of the $i^{\text{th}}-$atom, respectively, and $k_B$ is the Boltzmann constant. We assume that associated with each atom, there is a \emph{local electronic temperature}, i.e., $T_i^e$ that can fluctuate from atom to atom. {\color{black}Noteworthy, this local electronic temperature field is clustering the contribution of many electrons surrounding the atoms.} Let us now define the maximum $T^e_{\text{max}}$ and minimum $T^e_{\text{min}}$ allowed electronic temperatures in the system. Practically, $T^e_{\text{min}}= 0$ K while $T^e_{\text{max}}$ is selected sufficiently large such that the electronic temperature in the system does not exceeds this value. These values are kept constant during a simulation.  Introduce the normalized temperature field $\theta_i^e = \frac{T_i^e - T^e_{\text{min}}}{T^e_{\text{max}}-T^e_{\text{min}}}$, which renders all normalized temperatures  to the interval $\theta_i^e  \in [0,1]$. 


Now, we are interested in simulating heat transport in molecular dynamics simulations and include the effect of \emph{both} phonons and electrons. In order to simulate heat conduction of the electronic temperature, we follow the approach proposed by Ponga and Sun \cite{Ponga:2018} and modified it to consider coupling between electrons and phonons. Thus, we use the following modified \emph{master equation}

\begin{equation} \label{eq:HeatTransport}
	\frac{\partial T_i^e}{\partial t} = T^e_{\text{max}} \sum_{{{\substack{ j =1 \\ j \neq i}}}}^{N_n} K_{ij} \lbrace \theta_j^e (1-\theta_i^e) \exp [\Delta e_{ji} ]  - \theta_i^e (1-\theta_j^e) \exp [ \Delta e_{ij}]  \rbrace - \frac{G}{C_e} (T_i^e - T_i^{\text{lat}}),
\end{equation}
where $K_{ij}$ is a pair-wise exchange rate thermal coefficient for the electronic temperature $[K_{ij} = \text{time}^{-1}]$, and $\Delta e_{ij} = -(\theta_i^e-\theta_j^e)$ is the normalized difference in the electronic temperature between the $i^{\text{th}}$ and $j^{\text{th}}$ atoms, and $C_e$ is the electronic heat capacity $[C_e = \text{energy} \cdot \text{tempearture}^{-1} \cdot \text{length}^{-3}]$. {\color{black} In our implementation, we use values of $K_{ij}$ and $C_e$ that are function of the temperature of the system.} {\color{black}Temperature dependence of $C_e$, and $K_{ij}$ are calculated using simple linear relation of  $C_e = \lambda T_e$, and $\kappa_e =\frac{\kappa_0T^e}{T^{lat}}$ where $\kappa_0$ is thermal conductivity at 273 K \cite{hoh:2000}. At each timestep $C_e$ and $K_{ij}$ are calculated for mean electronic and lattice temperature of the system.} 

The first term on the right hand side of Eq. \ref{eq:HeatTransport} measures the rate of electronic energy exchanged between two adjacent particles and it is arbitrarily chosen to be around the nearest neighbors of the $i^{\text{th}}$ particle, which makes the model fully local. However, it should be noted that extra neighbors can be selected and thus, rendering the model to a semi-local description. According to Ponga and Sun \cite{Ponga:2018}, the pair-wise exchange rate thermal coefficient can be computed from an asymptotic expansion of the master equation into the continuum limit leading to a relation with the (macroscopic) electronic thermal conductivity ($\kappa_e$) as

\begin{equation} \label{eq:Kij}
	K_{ij} = \frac{2\kappa_ed}{C_eZb^2},
\end{equation}
where $d$ is the dimension of the system, $Z$ is the coordination number, $b$ is the Burgers vector of the material.

The second term on the right hand side of Eq. \ref{eq:HeatTransport} is a \emph{linear} coupling term between the electrons and phonons. $G$ is a constant that defines the strength of the coupling between the atoms and the electrons and has units of $[G = \text{energy} \cdot \text{length}^{-3} \cdot \text{temperature}^{-1} \cdot \text{time}^{-1}]$ and $(T_i^e - T_i^{\text{lat}})$ is the temperature difference between the electronic and the lattice temperature of the $i^{\text{th}}$ particle. $T_i^{\text{lat}}$ denotes the lattice temperature of the $i^{\text{th}}-$atom defined as

\begin{equation}
	\textcolor{black}{T_i^{\text{lat}} =\displaystyle \sum_{j=1}^{N_j} \frac{\frac{1}{2} m_j {\bf v}_j\cdot{\bf v}_j}{\frac{3}{2} k_B N_j}}
\end{equation}
where ${\bf v}_j$ denotes the velocity of the $j^{\text{th}}-$atom with the velocity of the center of mass subtracted. The $j^{\text{th}}-$atoms are selected to be between a cut-off radius, i.e.,  $j \in {\bf x}_j - {\bf x}_i \le r_c$. In our work, the cut-off radius $r_c$ is arbitrarily selected to be the cut-off of the potential function. The total number of particles in this cut-off radius is denoted by $N_j$.

After modeling heat conduction in the electronic system, we notice that the last term on the right hand side of Eq. \ref{eq:HeatTransport} is the amount of energy exchanged between electrons and phonons. This amount of energy needs to be introduced(removed) into(from) the lattice, and several alternatives exist to do this. There are several implementations of this term in the literature including the works of Duffy and Rutherford \cite{duf:2007}, Caro and Victoria \cite{Caro89}, Finnis et al. \cite{Fin91}, H{\"a}kkinen and Landman \cite{Hak93}, where the energy is inserted/removed by computing the amount of work of a random force used to simulate a thermostat. An alternative approach is by modifying the equation of motion to include a damping force, i.e., Ivanov and Zhigilei \cite{iva:2003}. For simplicity, we followed the implementation later appraoch. The amount of energy exchanged per unit of time and volume at the $i^{\text{th}}$ particle between the electrons and the lattice is $G(T_i^e - T_i^{\text{lat}})$. This flux of energy needs to be introduced (removed) in the lattice by modifying its kinetic energy. In order to do so, we introduce this energy flux by modifying the dynamics of the lattice, i.e.,

\begin{equation} \label{eq:Motion}
	m_i \dot{{\bf v}}_i = {\bf F}_i + {\bf F}_i^{\text{damping}} = {\bf F}_i +  \xi_i m_i {{\bf v}}_i
\end{equation}
%
%
where $\dot{{\bf v}}_i$ is acceleration of the $i^{\text{th}}-$atom, ${\bf F}_i$ is the force acting on the $i^{\text{th}}-$atom due to the interatomic interactions, and $\xi_i m_i {{\bf v}}_i $ is the \emph{damping} force that appears due to the lattice heating (cooling) coming from the electronic temperature. $\xi_i$ is a coupling coefficient of the $i^{\text{th}}-$atom whose strength needs to be computed at each time step. Let us now analyze the amount of energy exchange by the electronic temperature per unit of time. Letting $V_{\text{atom}} = \frac{V}{N}$ be the atomic volume, the energy exchange between the lattice and the $i^{\text{th}}-$atom is

\begin{equation} \label{eq:RateExchange}
	\frac{dE_i^{e \rightarrow \text{lat}}}{dt} = G V_{\text{atom}} (T_i^e - T_i^{\text{lat}}).
\end{equation}
The rate of work (power) done by the damping force is
\begin{equation} \label{eq:WorkDone}
	\textcolor{black}{W ={\bf v}_i \cdot {\bf F}_i^{\text{damping}}  =  \xi_i  m_i {\bf v}_i \cdot {\bf v}_i .}
\end{equation}
It is easy to see that $\xi_i$ has to be computed at each time step in order to conserve energy in the system due to the exchange of energy between the electrons and the phonons. The rate work done by the damping force has to be equal to the amount of energy exchange by the electrons in Eq. \ref{eq:RateExchange}. leading to

\begin{equation} \label{eq:Xi}
	\textcolor{black}{\xi_i = \frac{ G V_{\text{atom}} (T_i^e - T_i^{\text{lat}})}{2K_i},}
	\end{equation}
where $K_i = \frac{1}{2}m_i {\bf v}_i \cdot {\bf v}_i$ is the local kinetic energy of $i^{\text{th}}-$atom. We notice that the quantity $K_i$ could be instantaneously zero at certain time steps. This would lead to instabilities in the simulation, as the coupling coefficient will tend to infinity. This is due to the fact that the work done by the damping force tends to zero, and thus, even if $\xi_i$ is large, it cannot accommodate this situations which ultimately affect the stability of the system. In order to avoid these numerical instabilities, we opted for avoiding energy exchange between phonons and electrons when $K_i < \epsilon$ at the $i^{\text{th}}-$atom, where $\epsilon$ is a tolerance of $10^{-5}$ eV. This simple solution works well and does not affect the thermal behavior of the system.

Before closing this section, we make some remarks. First, we notice that the formulation uses local electronic and lattice temperature, and therefore, the energy exchange happens locally in the atoms. The strength of the coupling between electrons and phonons, given by $\xi_i$ changes atom by atom and it is a function of the time. Hence, in the current form, the electronic temperature serves as a \emph{local} thermostat that tends to equilibrate the electronic and lattice temperatures.

\subsection{Implementation} \label{Implementation}
We have implemented the proposed ${\ell}$2T-MD model in the LAMMPS code. We have done so by including an electronic temperature variable per atom. Then, the master equation Eq. \ref{eq:HeatTransport} was implemented using a fix operation, which is called every integration step to evaluate the electronic temperature. The electronic temperature for each atom is calculated within a fixed cut-off radius that can be given by the user in the script. Once the electronic temperature is updated, the coupling strength between phonons and electrons (Eq. \ref{eq:Xi}) is computed and the equation of motion (Eq. \ref{eq:Motion}) is modified to include the damping force at the $i^{\text{th}}-$ atom. The fix was implemented to work for parallel calculations using the Message Passing Interface (MPI) library. Since the electrons thermal diffusivity is very high, the time step for the integration of Eq. \ref{eq:HeatTransport} has to be a fraction of the integration time step in MD simulations ($\sim$ 1 fs). Thus, during a lattice integration step, we actually allow to perform multiple integration steps of the heat equation, and this can be selected by the user from the simulation script. Additionally, in large energy cascade simulations ($>$ 10 KeV), the velocity of the ions could be extremely high, and thus sometimes it is desirable to have a variable integration time step for the lattice. Our implementation is capable of handling this if the user specifies it. Cascade simulations described in Section \ref{Section4} were carried out using a variable time step. The source code can be found in its online site \cite{l2T-MD:website}.

\section{Heat Conduction in a Ni bar}

\subsection{Finite Difference Method}
{\color{black}Our first test consists of a classical two-temperature approach, where lattice and electronic temperatures are modeled. Our goal is to show the ability of the master equation approach to solve diffusion equation in a system of particles, and to compare it with the classical Fourier law. In order to do so, we solve a heat conduction problem in a one dimensional (1D) bar. In the classical two-temperature approach, a Fourier law is used to model both electron and phonon transport, i.e, 

\begin{eqnarray}\label{eq:1d_FE}
 \begin{cases} 
		 C_e\frac{\partial T^e}{\partial t} = \kappa_e\frac{\partial^2 T^e}{\partial x^2} - G(T^e - T^{\text{lat}}) \label{eg:1d_FE_e}\\
		 \rho C_{\text{lat}}\frac{\partial T^{\text{lat}}}{\partial t} = \kappa_{\text{lat}}\frac{\partial^2 T^{\text{lat}}}{\partial x^2} + G(T^e - T^{\text{lat}}) \label{eg:1d_FE_l}.
 \end{cases} 
\end{eqnarray}

Our test consists in replacing the Fourier law for electrons transport by the master equation while keeping the Fourier law for the lattice heat conduction. Thus,  Eq. \ref{eg:1d_FE_e}(a) is replaced by the master equation (Eq. \ref{eq:HeatTransport}). The equations were solved using a custom code in MATLAB. The mesh points coincided with the atomic positions, which are separated a distance $b = \sqrt{2}a_0/2$, where $a_0$ is the lattice parameter. The evolution of the lattice temperature was carried out using the FD method. In order to evaluate the Laplace operator of the temperature, we used a three-point FD stencil, and the temperature was integrated using a Euler forward algorithm. Periodic boundary conditions were enforced along the $x$-axis for both electrons and phonons. 

\begin{figure}[!ht]
	\centering\captionsetup[subfloat]{labelfont=bf}
	\subfloat[Fourier (lattice \& electronic)]{\label{fig:1d_me} \includegraphics[width=0.5\textwidth]{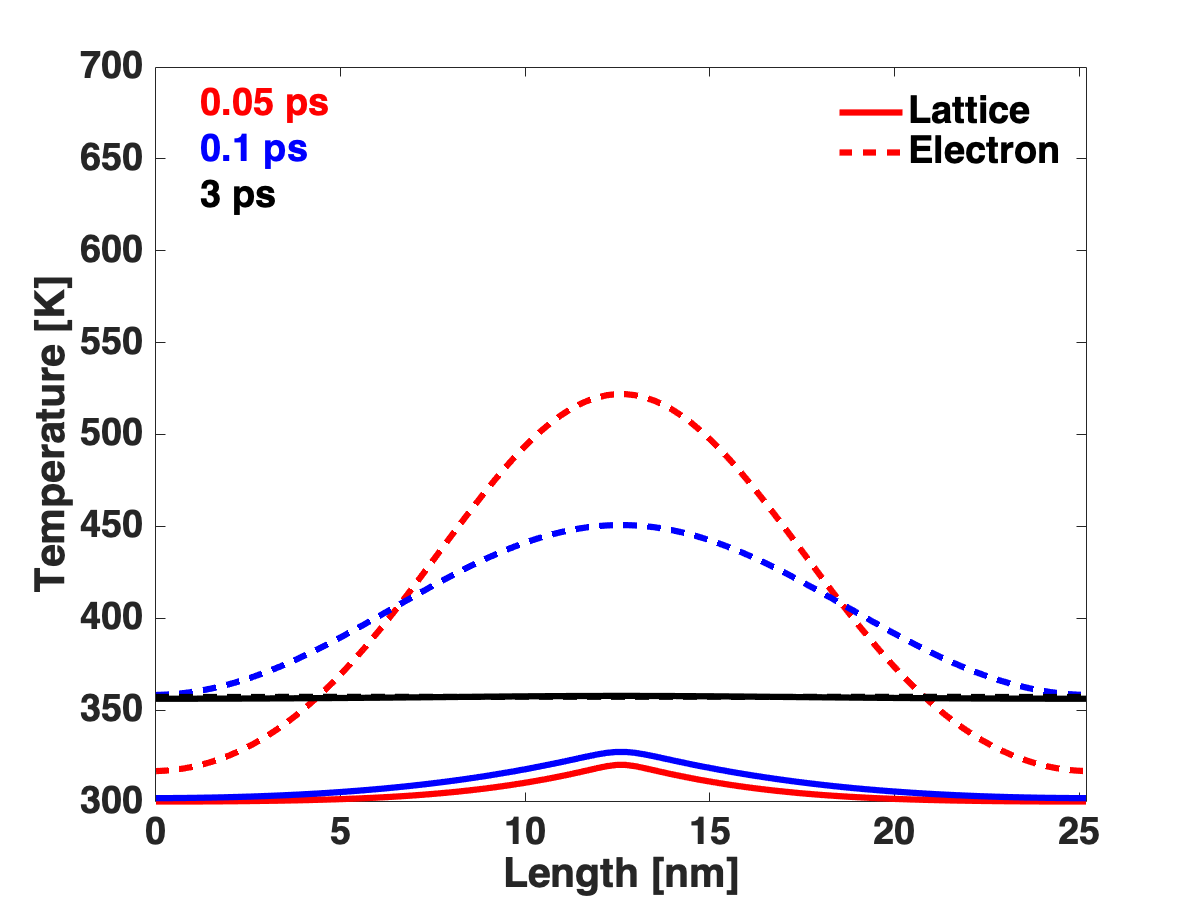}}
	\subfloat[Fourier \& Master Equation]{\label{fig:1d_md} \includegraphics[width=0.5\textwidth]{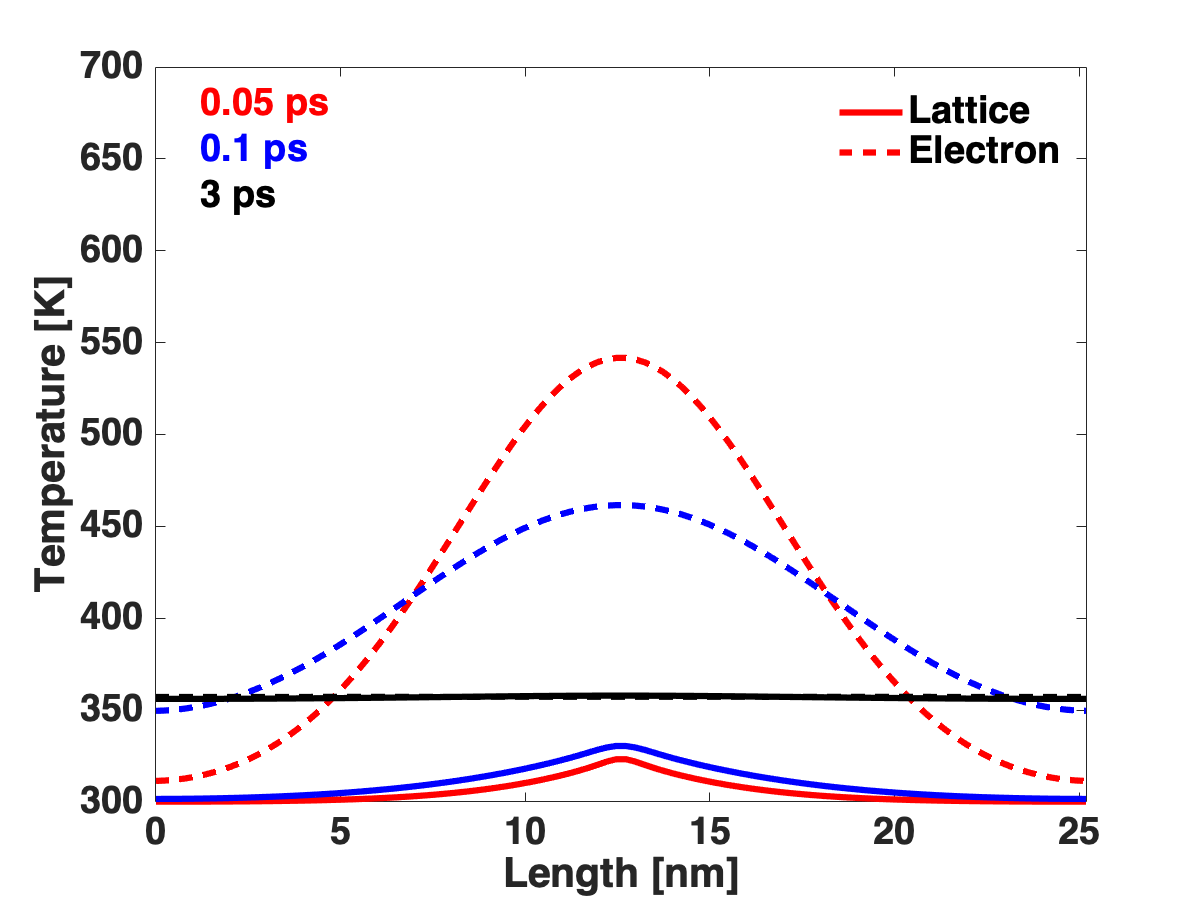}}
	\caption{(a) Time evolution of temperature alone the Ni-bar using the (a) Fourier law, and (b) master equation for the electronic temperature. Lattice temperature evolution was modeled with the Fourier law. The solid and dotted lines represent lattice and electronic temperature, respectively.}
\label{fig:1d_bar}
\end{figure}

Both lattice and electronic subsystems were kept at 300 K at the beginning of the simulation. A hot zone was created by applying 6000 K to the electronic subsystem to a span of length 2$a_0$ at the center of the bar. The dimensions of the Ni-bar was $l_x$ = 100$a_0$ = 25 nm, where $a_0$ = 0.3562 nm is the lattice constant of fcc Ni. In the FD model, electronic temperature dependence was not considered for heat capacity ($C_e$) and electronic thermal conductivity ($\kappa_e$). The material constants used for the description of the electronic subsystem are, electron-phonon coupling ($G$) = 3.6 $\times$ 10$^{17}$ J m$^{-3}$ K$^{-1}$ s$^{-1}$, electronic heat capacity ($C_e$) = 3.6886 $\times$ 10$^5$ J m$^{-3}$K$^{-1}$ and thermal conductivity ($\kappa_e$) = 91 J s$^{-1}$ m$^{-1}$  K$^{-1}$ \cite{iva:2003, hoh:2000} while for the lattice we took a fraction of it, specifically $\kappa_{\text{lat}}= 0.05 \kappa_e$\footnote{In metallic materials, electronic thermal conductivity is much larger than the lattice one, and thus the reason for this assumption.}. These parameters are used to simulate Ni. The pair-wise exchange rate thermal coefficient $K_{ij}$ was calculated using Eq. \ref{eq:Kij} with $d$ = 1 and $Z$ = 2. The lattice heat capacity ($C_l$) = 400 J Kg$^{-1}$  K$^{-1}$ and density ($\rho$) = 8890 Kg m$^{-3}$ were used for the lattice heat diffusion. A constant time step of 0.1 fs was used for both simulations.

The temperature profiles of the classical approach and proposed master equation model are shown in Fig. \ref{fig:1d_bar} (a) and (b), respectively. In the figure, we observe how the heat is being diffused following a Gaussian profile for the electrons, as predicted by the analytical solution. The large thermal diffusivity of the electrons spreads the high temperature very quickly, as can be seen in the graphical representation. Since the lattice thermal diffusivity is much smaller than the electrons, the rise of the lattice temperature is less evident and concentrated near the hot spot. We also notice that after 3 ps, both lattice and electronic temperature are in equilibrium. We also see that there is a very good agreement between the two models and only a small differences appear in the electronic temperature. These differences can be due to differences in the thermal conductivity in the models, as the pair-wise exchange rate given in Eq. \ref{eq:Kij} is obtained using a long wave analysis, and some small differences can be seen in this case, where the heat is being dissipated very fast. Besides these differences, we see that the equilibrium temperature at the end of the simulation is the same, an indication that the energy is conserved in the system. Next, we proceed to introduce electronic heat transport in MD simulations in the next example. }

\subsection{Molecular dynamics simulations}

\begin{figure}[!ht]
	\centering 
	\label{fig:etemp_1}\includegraphics[width=0.7\textwidth]{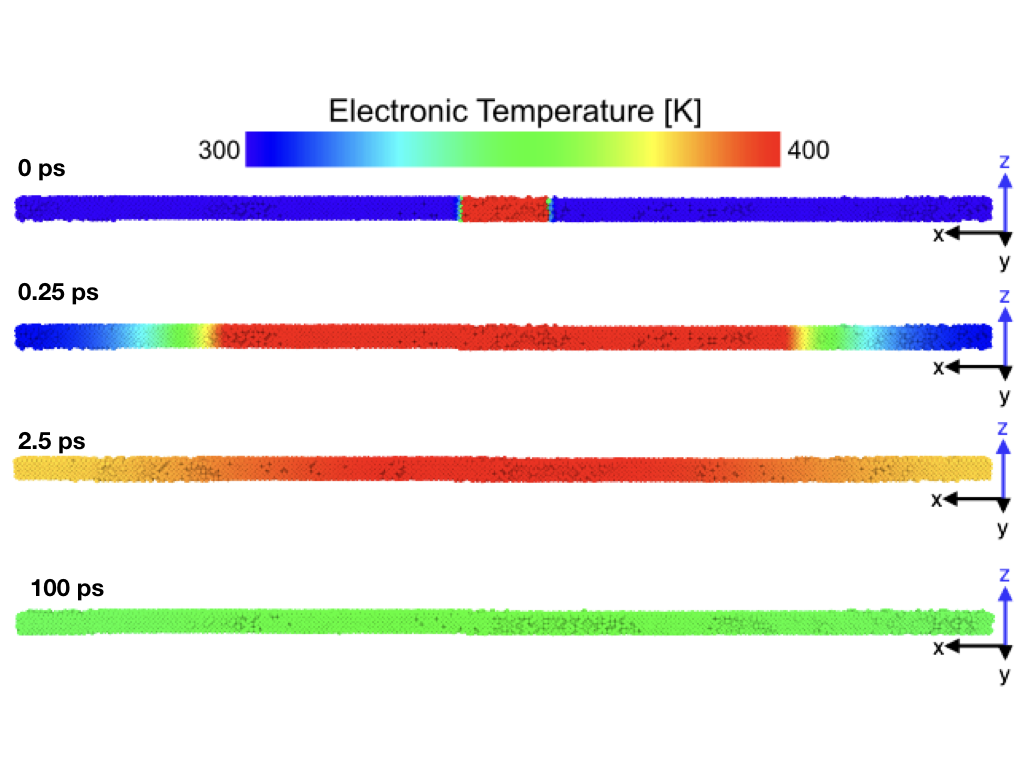}
%
	\caption{Snapshots of the electronic temperature field in the 1d-Ni bar using the ${\ell}$2T-MD method at different time steps.}
	\label{etemp_1d_1_3}
\end{figure}

{In this section, we describe a simple ${\ell}$2T-MD model which is analogous to the previous one. We took a Ni-bar of dimensions $l_x$ = 25 nm -same as model above- and $l_y$ = $l_z$ = 2$a_0$ = 0.71 nm. The $C_e$ and $K_{ij}$ were function of electronic temperature ($T_e$) as described in Section \ref{Section:Methodology}, Eq. \ref{eq:Kij}, using $\lambda$ = 1065 J m$^{-3}$K$^{-2}$, ($\kappa_0$) = 91 J s$^{-1}$ m$^{-1}$  K$^{-1}$, $d$ = 3 and $Z$ = 12. The integration time step of MD was 0.5 fs while the electronic one was 10 less. Thus, for each MD time step, we performed 10 electronic heat conduction steps. Notice that even though the heat is being dissipated in one direction, our example is fully three-dimensional and no simplifications are made for the electronic heat diffusion \footnote{One could, in principle, input the value of $K_{ij}$ using Z = 2 and d=1 -as in the one dimensional case- to obtain the same equilibration time as in the previous example. We have done so and the lattice and electronic temperature reach equilibrium about 3.2 ps, in a good agreement with the previous example.}. Fig. \ref{fig:1d_bar_md} shows the time evolution of the electronic and lattice temperature in the MD sample. The lattice temperature was computed by dividing the computational cell in individual bins along the $x-$direction, and computing the temperature of the atoms in that bin ($N_b$) for 20 time steps using the classical expression $T_{\text{lat}} = \frac{2 K_e}{3 k_B N_b}$. The value of $T_{max} = 20,000$ K, sufficiently large to map the temperature field to the segment $[0,1]$. 

Figs. \ref{fig:1d_bar_md} (a)-(d) show the time evolution of the electronic and lattice temperature for $\Delta t = 1, 2, 10$, and $99.5$ ps, respectively. We see that the electronic temperature is quickly distributed across the sample, as depicted in Figs. \ref{fig:1d_bar_md} (a) and (b).  However, after $\Delta t = 10$ ps, there are some small differences between them due to poor lattice diffusivity, with a larger lattice temperature in the middle of the sample, where the initial hot spot was placed. After $\Delta t = 99.5$ ps, we observed that both are in equilibrium. Remarkably, once the system is in equilibrium, both electronic and lattice temperatures reach the same value, with some statistical fluctuations in the lattice temperature. The final mean temperature of both electronic and lattice subsystem reached equilibrium at $\sim$350 K.

\begin{figure}[!ht]
	\centering
	\subfloat[]{\label{fig:etemp_a}\includegraphics[width=0.45\textwidth]{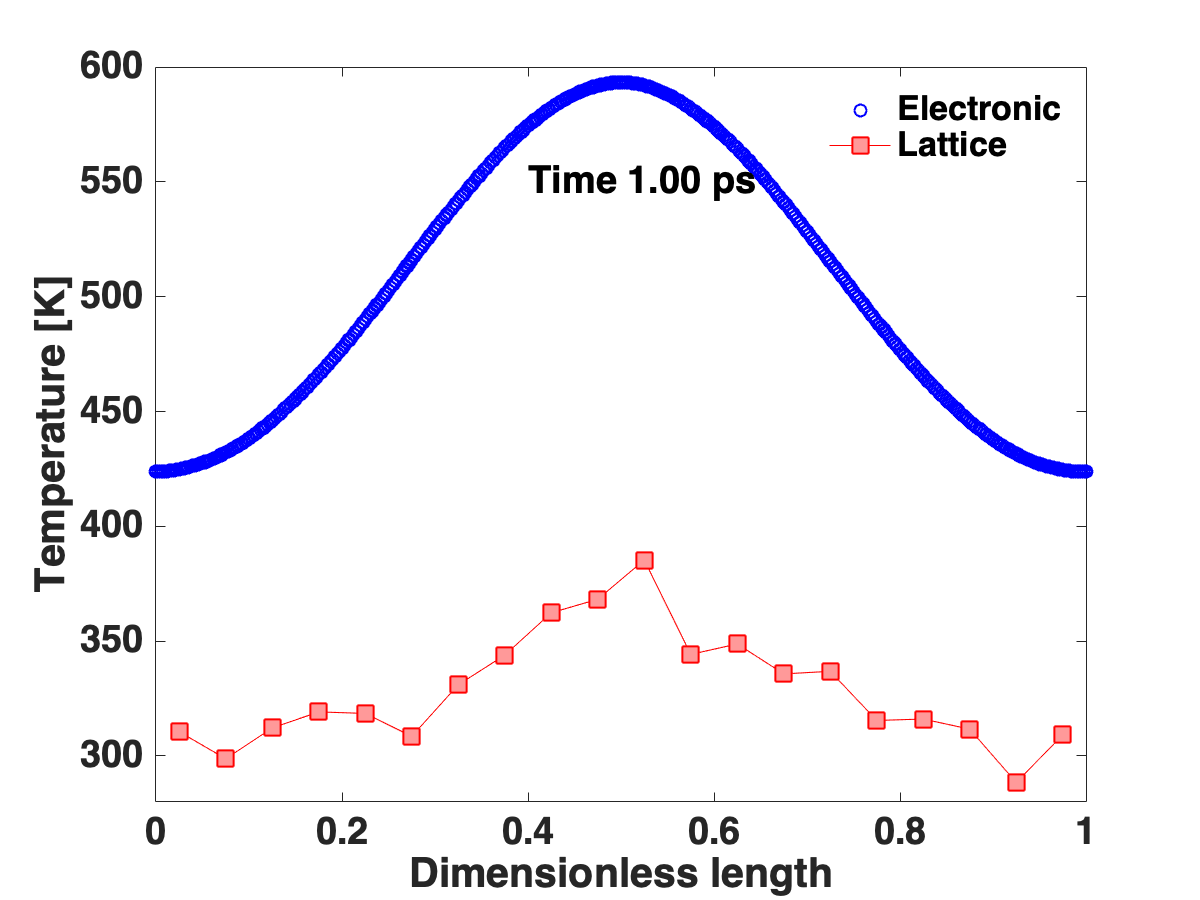}}
	\subfloat[]{\label{fig:etemp_b}\includegraphics[width=0.45\textwidth]{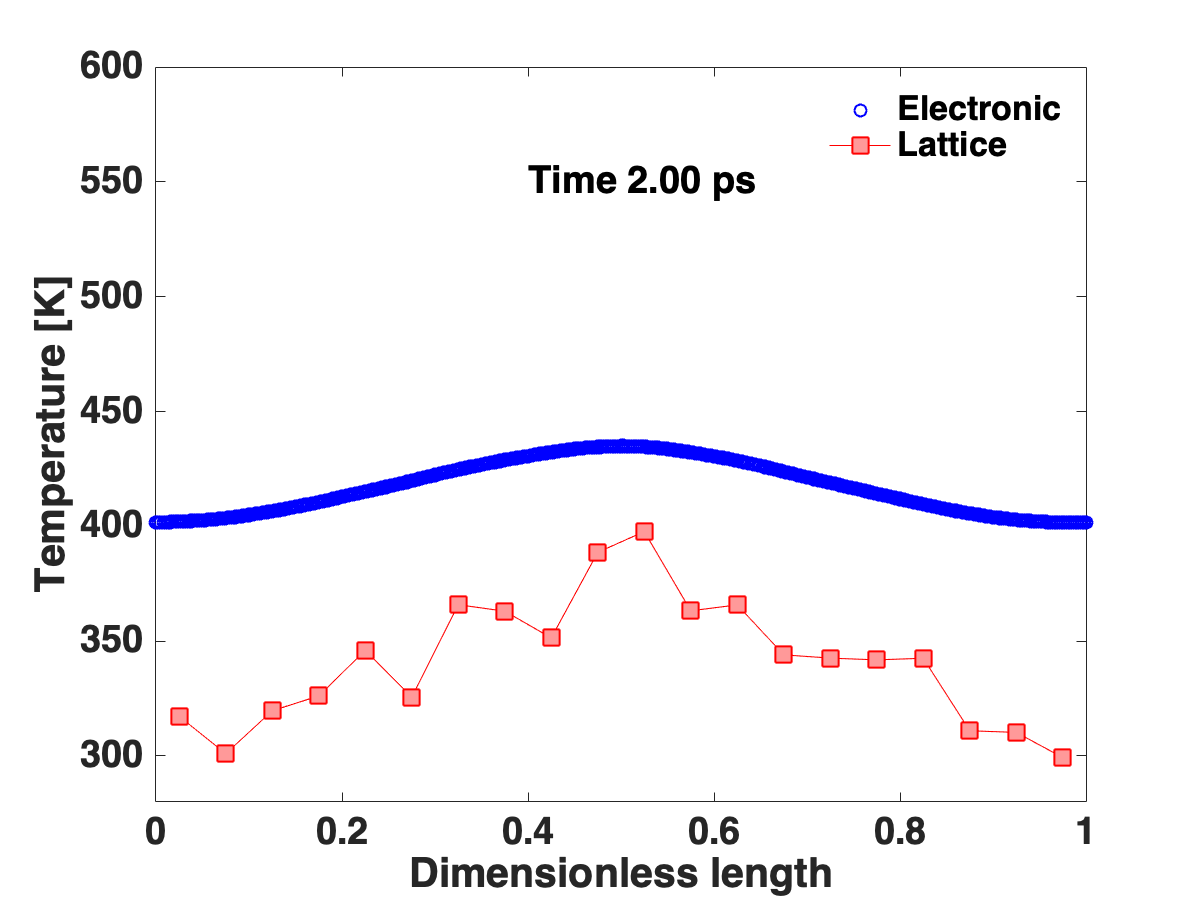}} \\
	\subfloat[]{\label{fig:etemp_c}\includegraphics[width=0.45\textwidth]{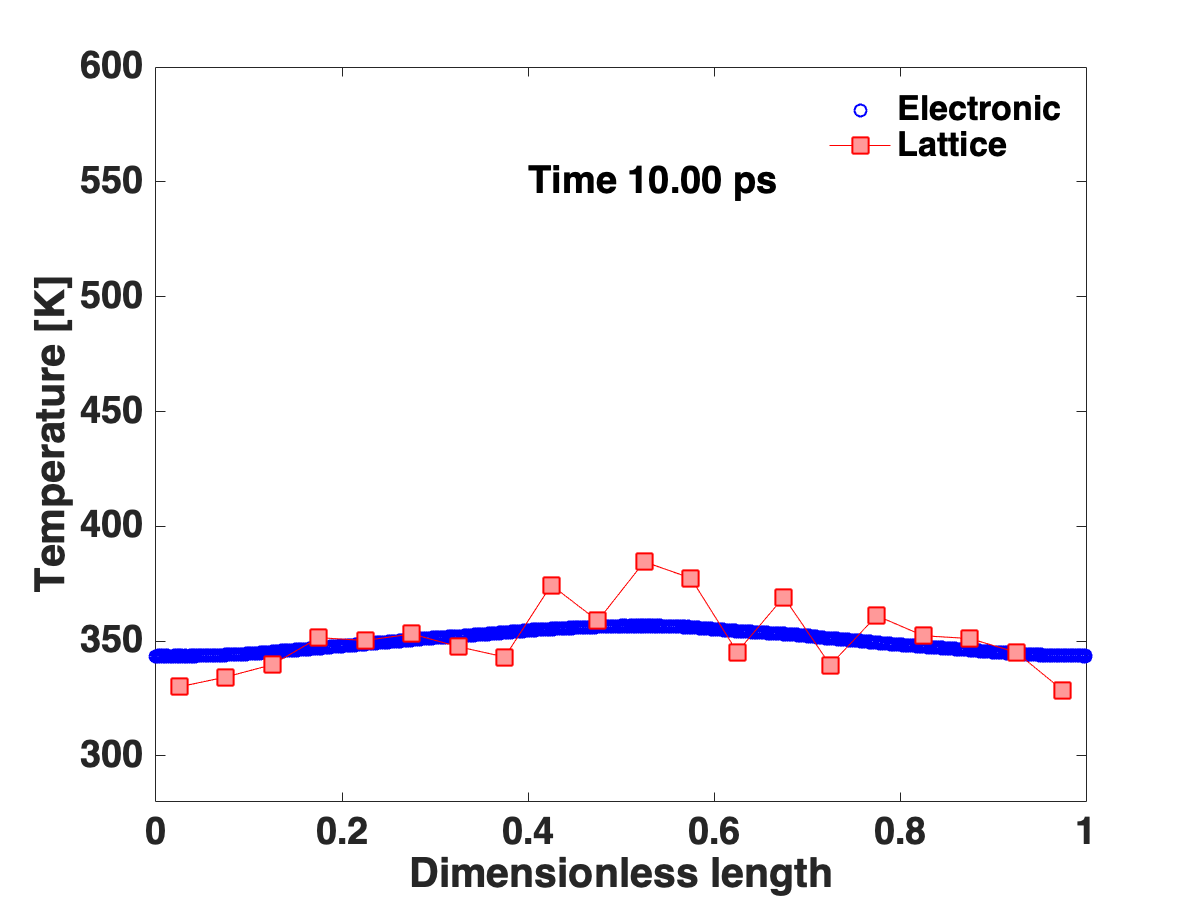}}
	\subfloat[]{\label{fig:etemp_d}\includegraphics[width=0.45\textwidth]{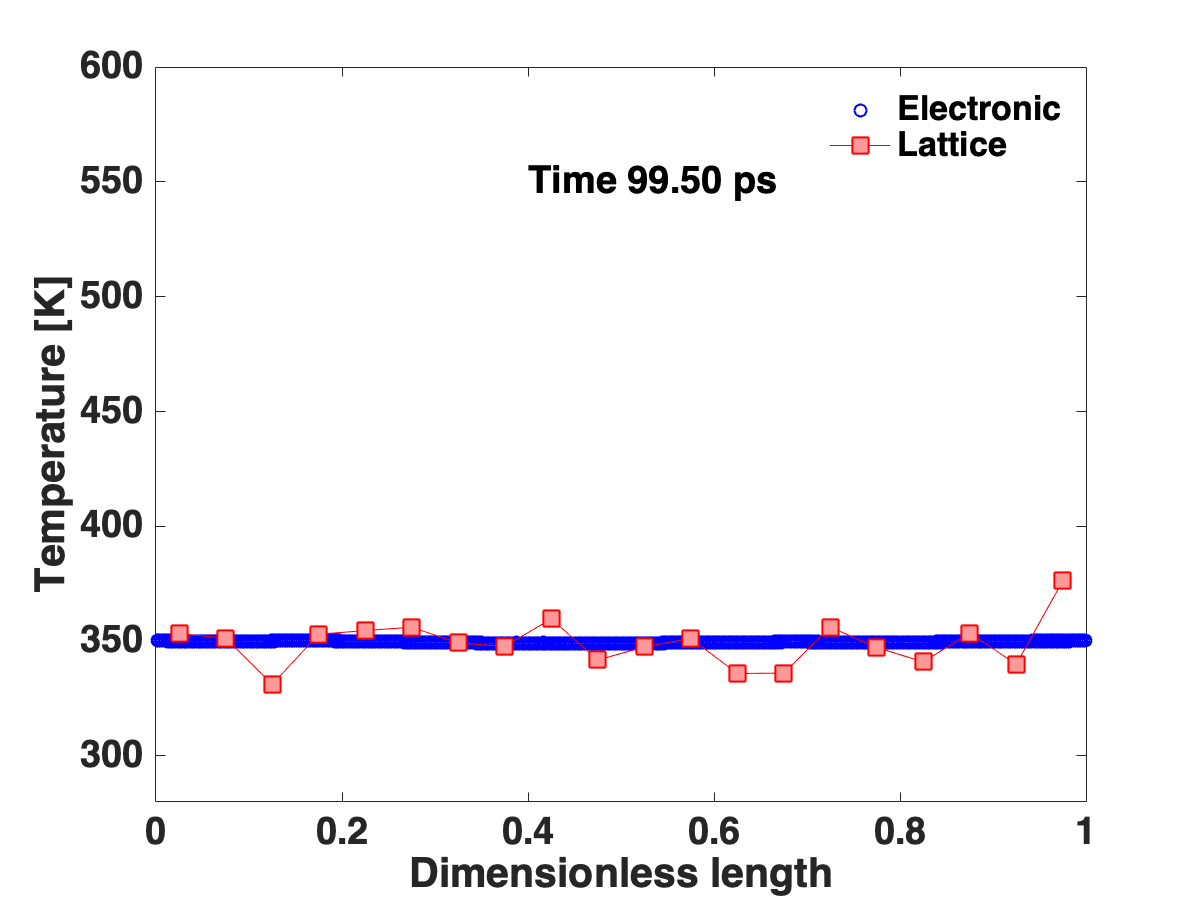}}
	\caption{Time evolution of the electronic (blue) and lattice (red) temperature alone the Ni-bar using MD simulation}
	\label{fig:1d_bar_md}
\end{figure}

Figure \ref{etemp_1d_1_3} shows the snapshots of electronic temperature distribution of ${\ell}$2T-MD simulation at different time steps. We see the ability of the model to retain an atomic resolution for the electronic temperature.}

\begin{figure}[!ht]
	\centering
	\includegraphics[width=0.60\textwidth]{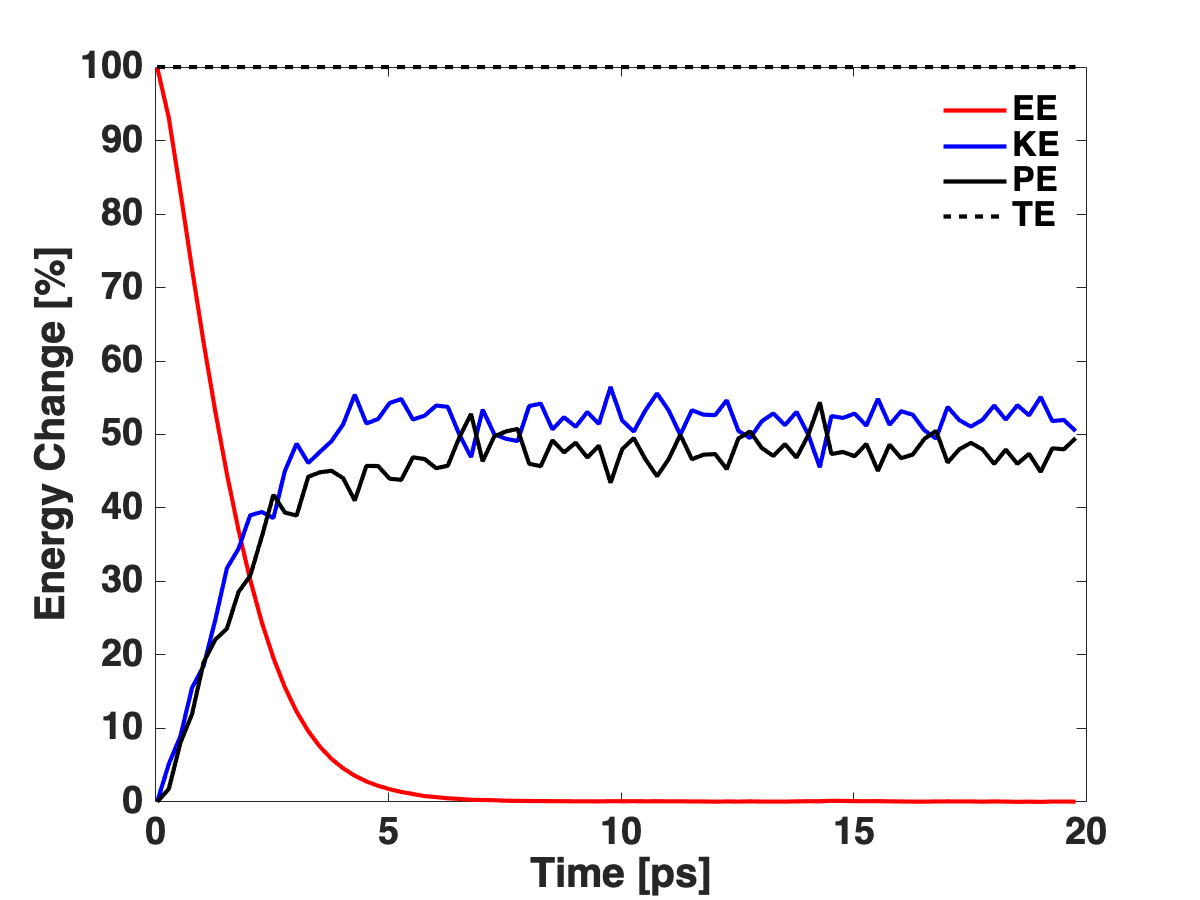}
	\caption{Energy Change \emph{vs.} time for the electronic (red), kinetic (blue) and potential (black) energies during a ${\ell}$2T-MD simulation for the uniaxial bar. Energies (E) are shifted by subtracting their corresponding minimum values (E = E - $min$(E)) then normalized to the total energy (E = TE/E). A six-degree polynomial fit is used to plot the kinetic and potential energies. The  normalized total energy of the system (TE = (EE + KE + PE)/TE) is plotted with black dashed lines.}
	\label{fig:1d_bar_energy}
\end{figure}

{The energy balance of the ${\ell}$2T-MD model is shown in Figure \ref{fig:1d_bar_energy}. The total energy (TE) is conserved in the simulation. Initially, the electronic energy was high as a high temperature was given to the electronic subsystem. The energy was gradually transferred to the atomic subsystem and increasing both kinetic and potential energies. We see that after $\sim$ 5 ps, the energy exchange reduces significantly and the three quantities remain approximately the same, as expected.} 

\begin{figure}[!ht]
	\centering\captionsetup[subfloat]{labelfont=bf}
	\label{fig:1d_FE}\includegraphics[width=0.45\textwidth]{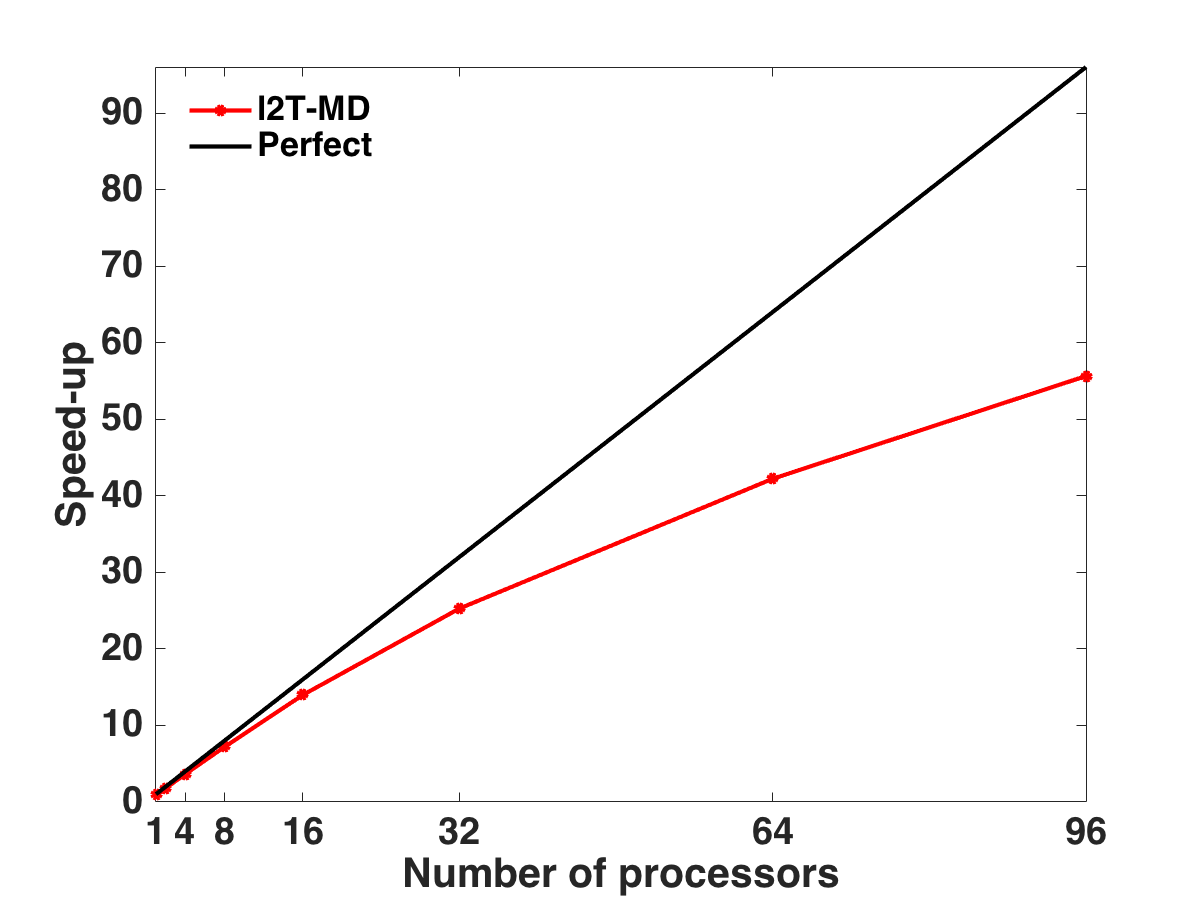}
	\caption{Scaling of the $\ell$2T-MD model implemented in LAMMPS. We see that the implementation retains about 80\% of the efficiency up to 32 processors. After this point, the communication is the bottleneck and it gets reduced to $\sim$60\% for 96 processors.}
	\label{Speed-up}
\end{figure}

Figure \ref{Speed-up} shows the performance of the implementation in the LAMMPS code for the uniaxial bar problem with a larger number of atoms, i.e., $N=40,000$. As we can see, the implementation has good scalability up to 32 processors (80\% efficiency). However, after this point, the communication becomes the bottleneck and the efficiency gets reduced ($\sim$ 60\% for 96 processors). This is due to two main reasons, the number of atoms per processor reduces to less than 1000 atoms/processors, and therefore, it is difficult to scale since the ratio between computing over communication time is reduced significantly. Second, in order to the electronic heat diffusion, we need to communicate information to the ghost atoms in the simulation, which is an overhead with respect to MD. Thus, this reduces the performance. Reduction of communication can be achieved by using hybrid implementations with threads in OpenMP. This will be addressed in the future. 

\subsubsection{Heat conduction in a thin film}

Next, we proceed to illustrate the behavior of the system when the bar is changed to a plate. In this case, we selected a length $l_x = l_y = 50 a_0$ and $l_z = 4a_0$. Periodic boundary conditions are used in $x-$ and $y-$ directions and free surface is simulated in the $z-$direction. Initially, we ran a traditional MD simulation until the lattice temperatures was in equilibrium at $T=300$ K using the Canonical ensemble (NVT). The electronic temperature was also initialized to same value for all atoms. Then, an initial large value of the electronic temperature $T_e = 10,000$ K was given to all atoms in the region $6a_0 \times 6a_0$ in the center of the simulation cell. The material parameters were kept the same than the previous example.

\begin{figure}[!ht]
	\centering 
	\subfloat[]{\label{fig:plate1}\includegraphics[width=0.28\textwidth]{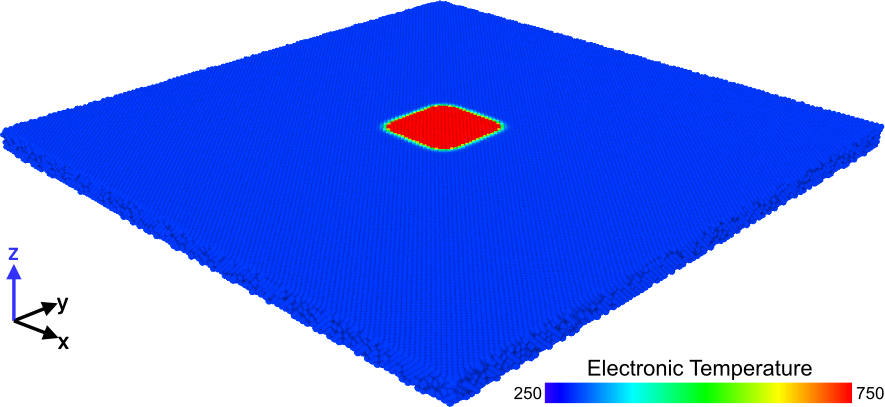}} \hspace{1cm}
	\subfloat[]{\label{fig:plate2}\includegraphics[width=0.28\textwidth]{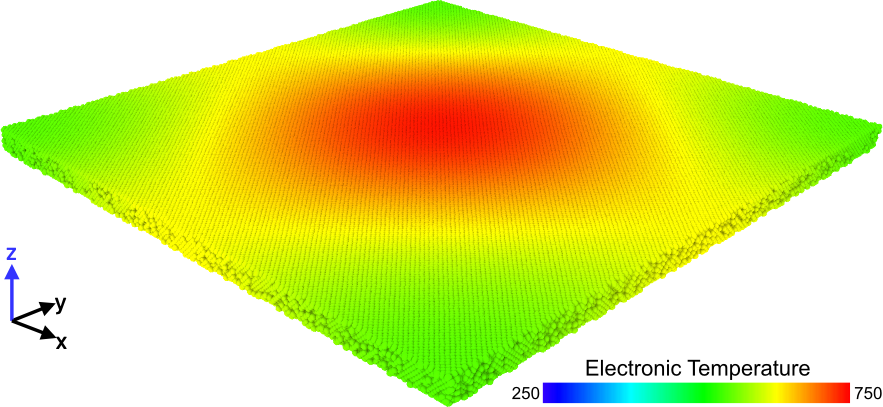}}  \hspace{1cm}
	\subfloat[]{\label{fig:plate3}\includegraphics[width=0.28\textwidth]{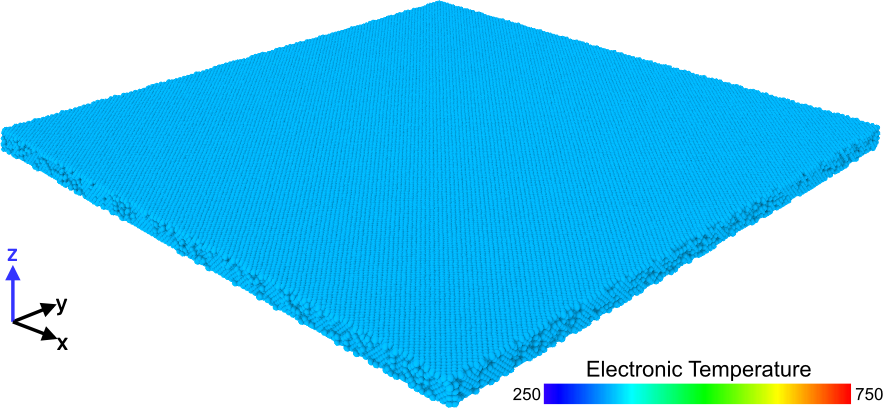}} \\
	\subfloat[]{\label{fig:plate1}\includegraphics[width=0.32\textwidth]{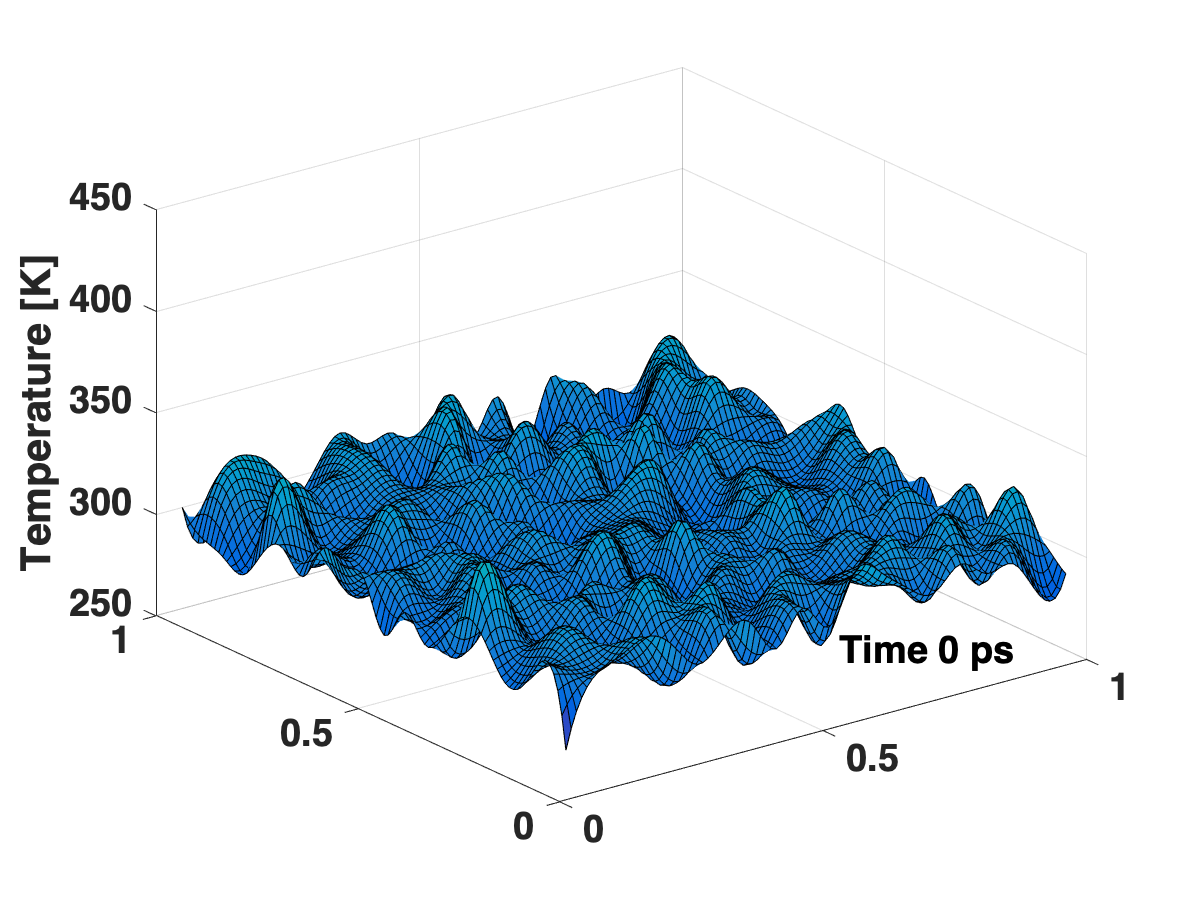}}
	\subfloat[]{\label{fig:plate2}\includegraphics[width=0.32\textwidth]{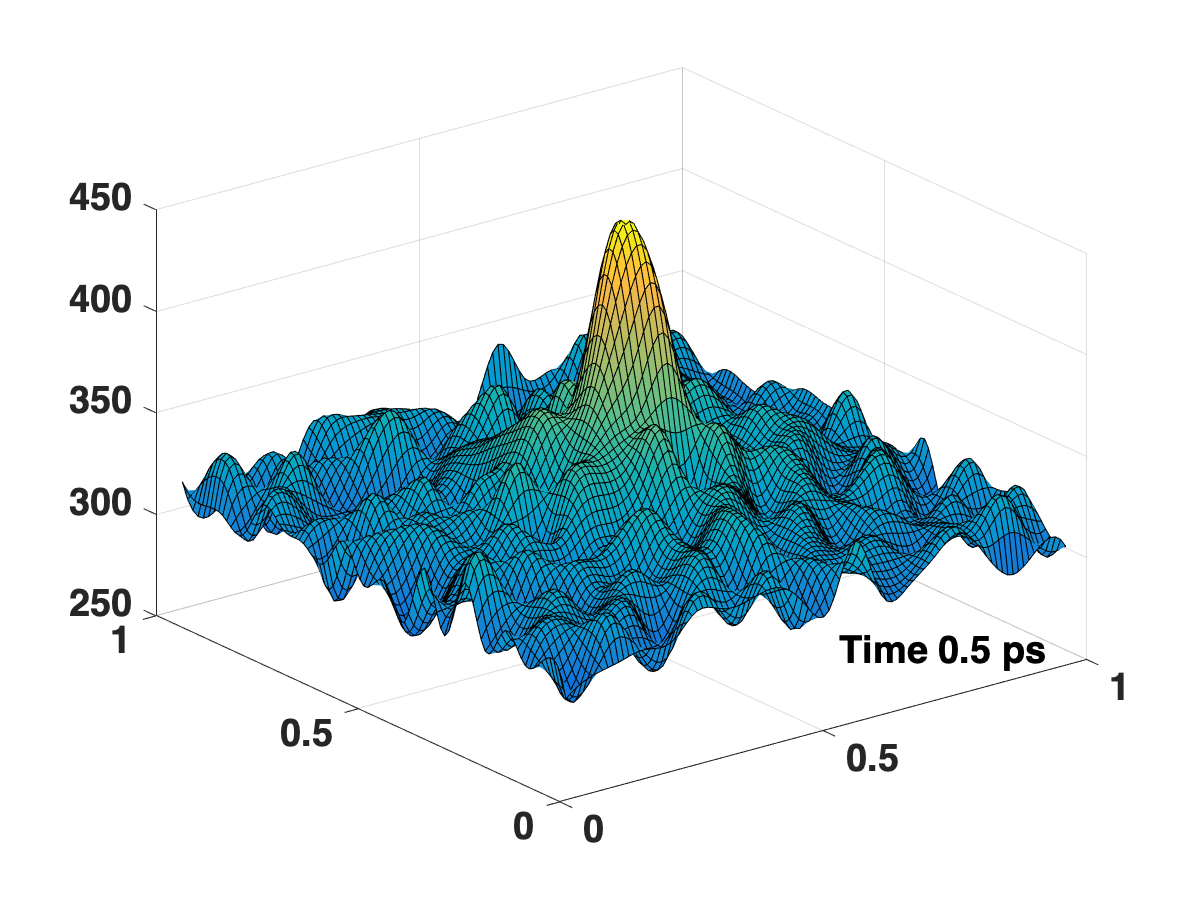}} 
	\subfloat[]{\label{fig:plate3}\includegraphics[width=0.32\textwidth]{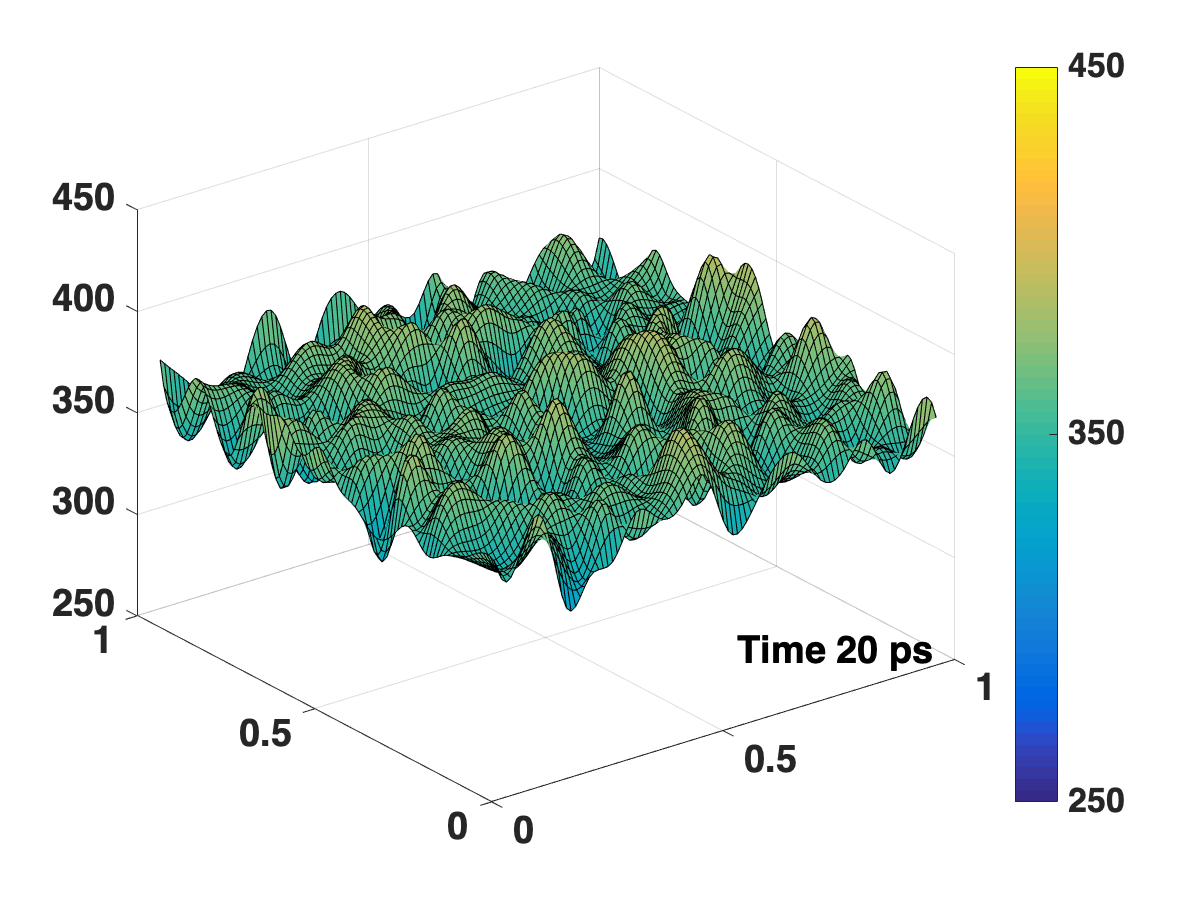}} 
	\caption{Snapshots of the electronic and lattice temperatures for the thin plate with periodic boundary conditions. (a)-(c) Electronic temperature; (d)-(f) lattice temperature.}
	\label{fig:plate}
\end{figure}

In order to monitor the lattice temperature, we have divided the simulation cell into $20 \times 20$ bins of equal size, and compute the lattice temperature every 0.5 ps for 20 time steps. With these values of the lattice temperature at each bin, we then performed a two-dimensional spline interpolation to produce a smooth surface for visualization purposes. Figure \ref{fig:plate} shows the time evolution of the electronic and lattice temperature in the simulation cell. Figs.  \ref{fig:plate}(a)-(c) show the electronic temperature for the atoms in the sample at different times. Clearly, the elevated temperature given to the central atoms is dissipated quickly across the sample, until the temperature is constant after 20 ps. On the other hand, the lattice temperature, shown in Figs. \ref{fig:plate}(d)-(f) with the surface plots, gets dispersed much slower than the electronic one, as can be seen in Figs. \ref{fig:plate}(e), where the central atoms are at about 450 K, to finally equilibrate with the electronic temperature at 20 ps at around 358 K. Remarkably, the fluctuations observed in the lattice temperature are characteristic of the statistic nature of the temperature.

\section{Irradiation damage modeling} \label{Section4}

Now, we turn our attention to simulate far from equilibrium phenomenon such as ion irradiation, where electronic excitations and ionizations are significant. Irradiation in metallic systems has been studied in several works including only the effect of the phonons \cite{Aidhy:2015,Ullah:2016}. Here, we include the effect of the electrons and phonons using the proposed approach. We performed 50 KeV Ni cascade simulations to model irradiation induced damage of elemental Ni using two different models (classical MD and ${\ell}$2T-MD). The cascade simulations were done using the newly implemented ${\ell}$2T-MD method in LAMMPS \cite{Lammps} code. We used the embedded atom method (EAM) based potential of Bonny \emph{et~al.} \cite{bonny:2013} to model elemental Ni, however, the ${\ell}$2T-MD method works with all potentials available in LAMMPS. This potential was developed to study the production and evolution of radiation defects, particularly, point defects. The simulation cell size was 34$\times$34$\times$34 nm$^3$ containing $\sim$3,500,000 atoms. The lattice and electronic subsystems were initially at equilibrium having a temperature of 300 K. Periodic boundary conditions were applied to all directions to mimic an infinitely large bulk system. \textcolor{black}{The phononic system was cooled down at the boundaries using Berendsen temperature control and a Dirichlet/fixed boundary condition was used for the electronic system to 300 K.} The electronic stopping was modeled as a friction force for atoms with kinetic energies higher than 10 eV and calculated using the Stopping and Range of Ions in Matter (SRIM) code version 2013 \cite{Nor94b, SRIM:2013}. A fixed time step of 1 femtosecond (fs) was used to relax the system and a variable time step was used having the maximum time step restricted to 0.01 fs for cascade simulations. At the beginning of the simulation, a recoil energy of 50 KeV was given to a central atom at a random direction. A total of 7 such simulations were performed to collect damage statistics. The simulation run at 300 K for 16 ps. 

\begin{figure}[!ht]
	\centering\captionsetup[subfloat]{labelfont=bf}
	\subfloat[]{\label{fig:max_etemp}\includegraphics[width=0.45\textwidth]{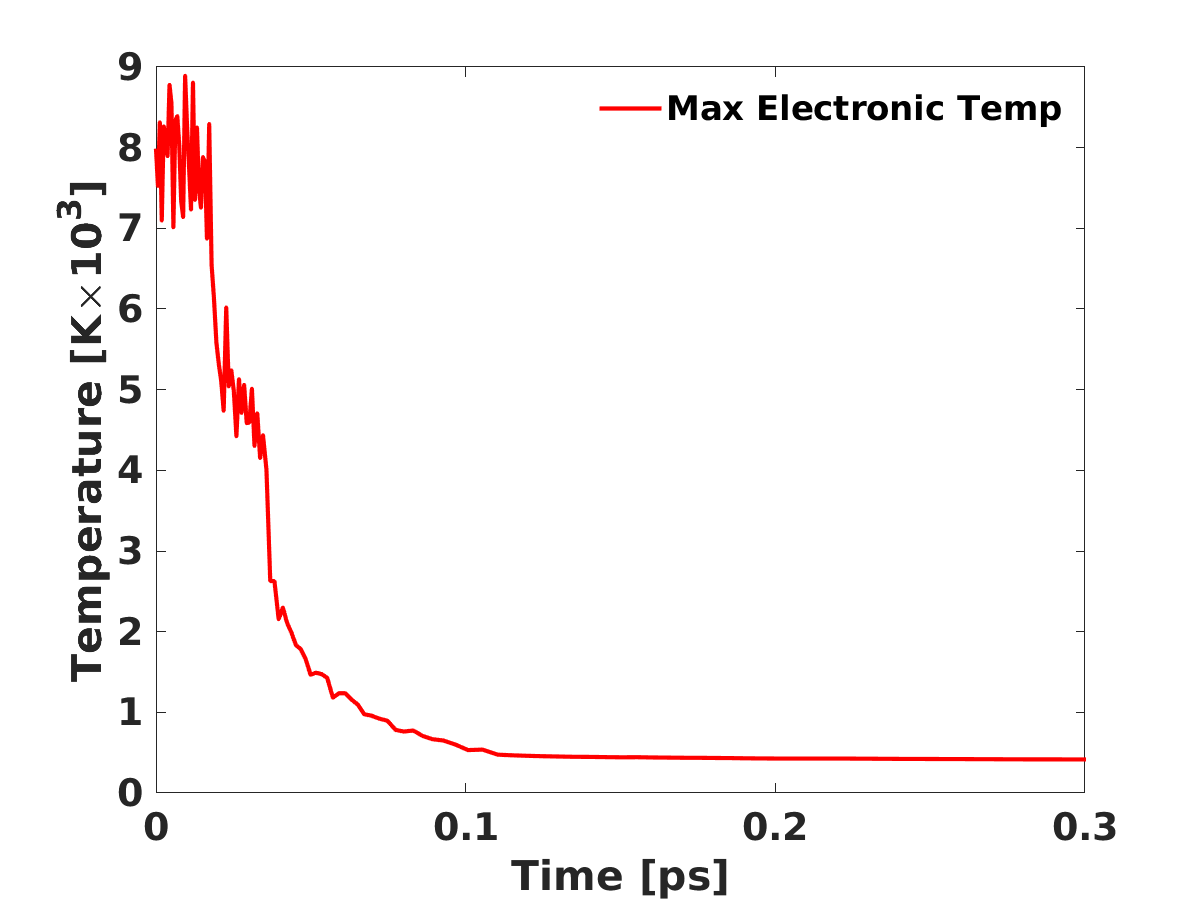}}
	\subfloat[]{\label{fig:max_ke}\includegraphics[width=0.45\textwidth]{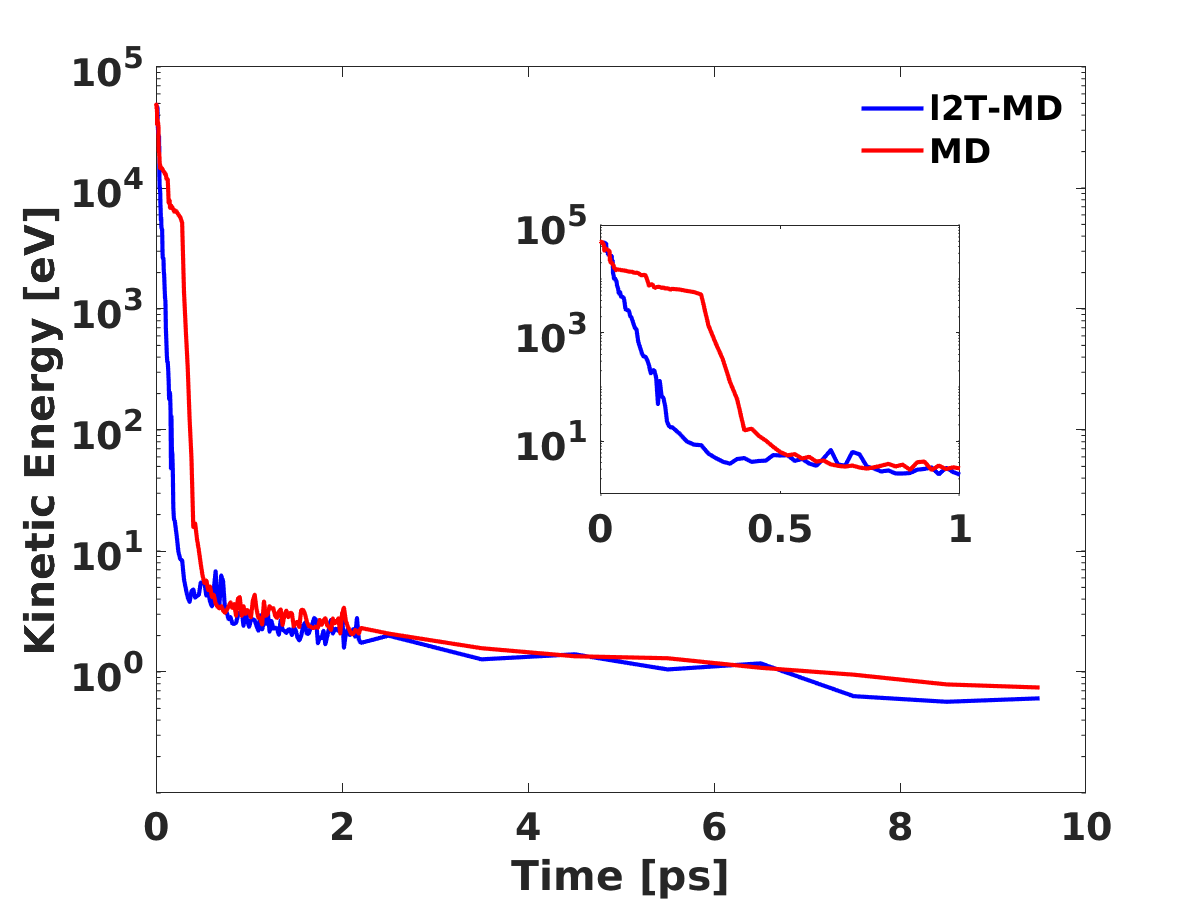}}
	\caption{ (a) maximum electronic temperature, and (b) maximum kinetic energy of a representative 50 KeV cascade simulation.}
	\label{mean_temp}
\end{figure}

The maximum electronic temperature is the highest atomic value calculated at the corresponding time step, which is shown in Fig. \ref{mean_temp}(a) as a function of time. It shows highly excited electrons in small regions that can reach temperatures in excess of 8000 K. We see that the maximum electronic temperature decreases very quickly and it reaches an equilibrium point at around 0.1 ps after the event. Now, let us turn our attention to the time evolution of the maximum lattice kinetic energy, shown in Fig. \ref{mean_temp}(b) where we have made comparison for regular MD and the $\ell$2T-MD model. Two main key aspects can be observed from the figure. First, the rate of change of the kinetic energy is much faster for the $\ell$2T-MD model, since the effect of electrons are taking into account and they contribute to dissipation. As a result, the maximum kinetic energy in the sample for the regular MD model remains large ($\sim$15 KeV) for approximately 0.5 ps, while in the $\ell$2T-MD model, the maximum kinetic energy is reduced immediately. For instance, the kinetic energy in the $\ell$2T-MD model is reduced from $\sim$35 KeV to $\sim$1 eV in around 2 ps while in the regular MD model this takes 3 ps. This affects the generation of defects, as we will explain later on. Second, the plot shows that local atoms in the sample are heavily excited as indicated by their kinetic energy and thus, are far from equilibrium in both MD and the $\ell$2T-MD models. 


\begin{figure}[!ht]
	\centering\captionsetup[subfloat]{labelfont=bf}
	\subfloat[]{\label{fig:temp_snap}\includegraphics[width=0.4\textwidth]{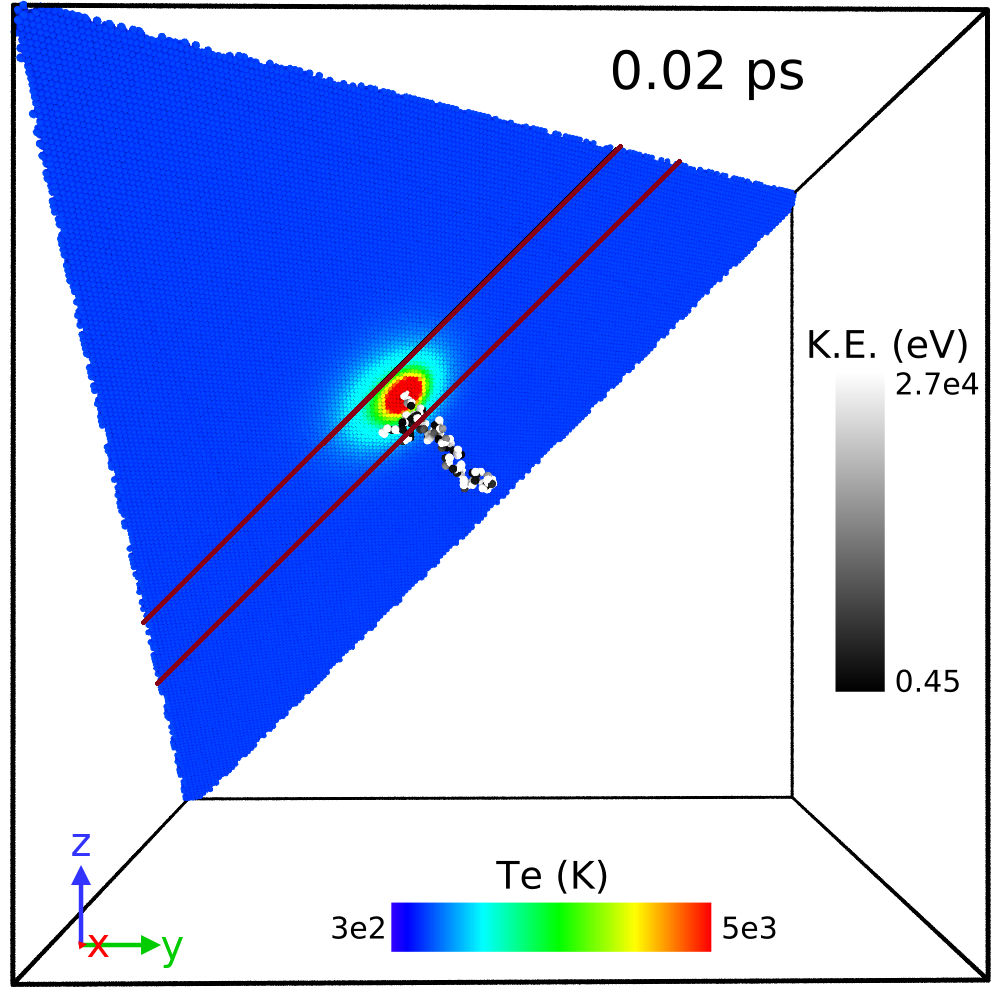}}
	\hspace{1em}
	\subfloat[]{\label{fig:temp_dist}\includegraphics[width=0.5\textwidth]{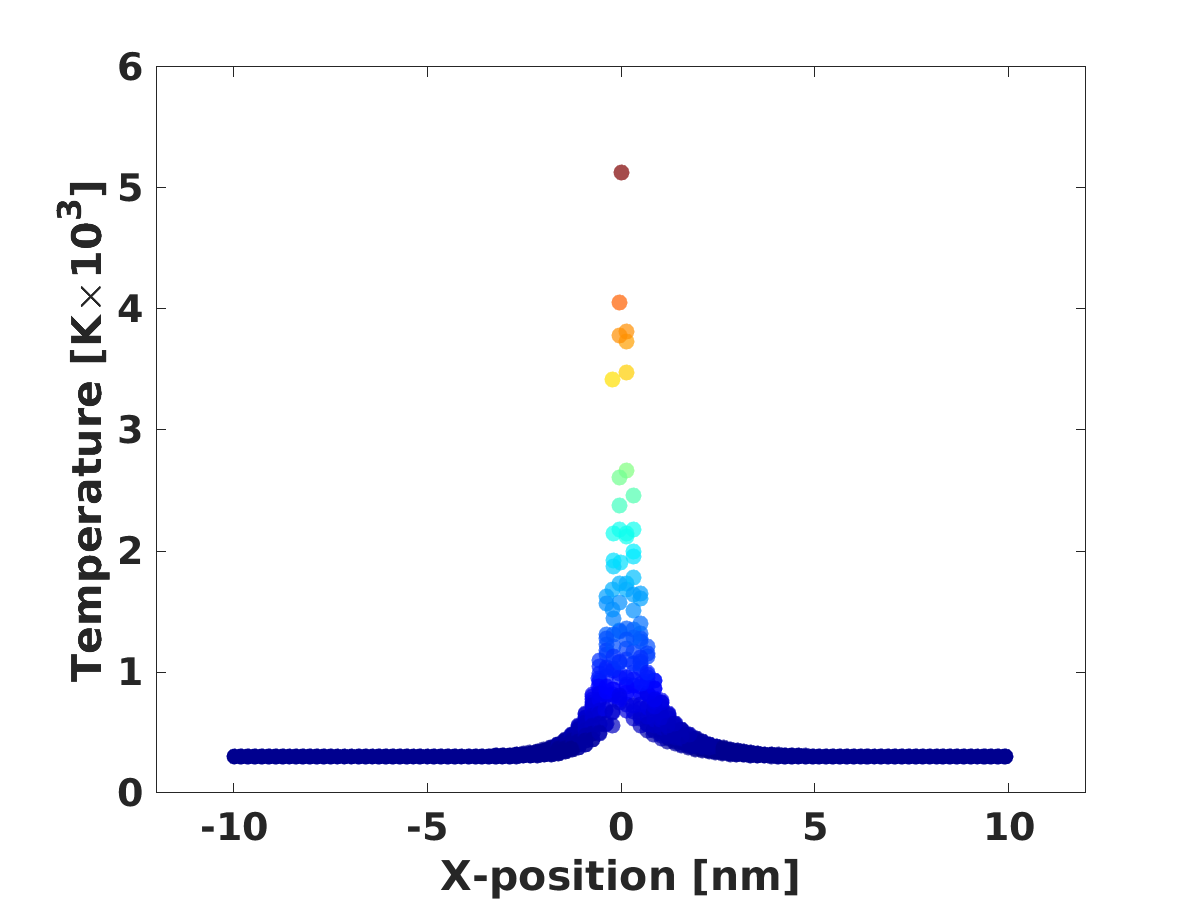}}
	\caption{(a) Snapshot of simulation cell at 0.02 ps after the recoil event. The gray color atoms are the energetic particles along the recoil path. The triangular plane shows the electronic temperature distribution at the front. (b) The electronic temperature distribution along the thin slice (brown lines on the triangular plane). Zero (0) is the center (hot spot) of the thin slice.}
	\label{etemp_dist}
\end{figure}

Figure \ref{etemp_dist}, shows the electronic temperature distribution of the 50 KeV recoil event after 0.02 ps together with atoms showing exceeding amounts of kinetic energy. We see that only a small fraction of atoms have kinetic energy in excess of 0.4 eV. However, the electronic temperature can reach exceedingly large values as shown in the color map. The rainbow color map shows a radial distribution of the electronic temperature on the plane, characteristic of a Gaussian type distribution. To have a better understanding of the temperature distribution, Fig. \ref{etemp_dist}(b) shows the temperature distribution along the narrow slit (brown lines) on the plane shown in  Fig. \ref{etemp_dist}(a). We observed a large temperature gradient ($\sim$ 250 K$\cdot$nm$^{-1}$) from the central hot zone to the surrounding. This is a clear indication of a large temperature gradient happening in a very small length scale, denoting the local character of the energy exchange in the system. Our analysis illustrates the need for models that are able to have a fine resolution in order to properly predict the electronic effects in these situations.

\begin{figure}[!ht]
	\centering\captionsetup[subfloat]{labelfont=bf}
	\subfloat[]{\label{fig:ttm_fp}\includegraphics[width=0.4\textwidth]{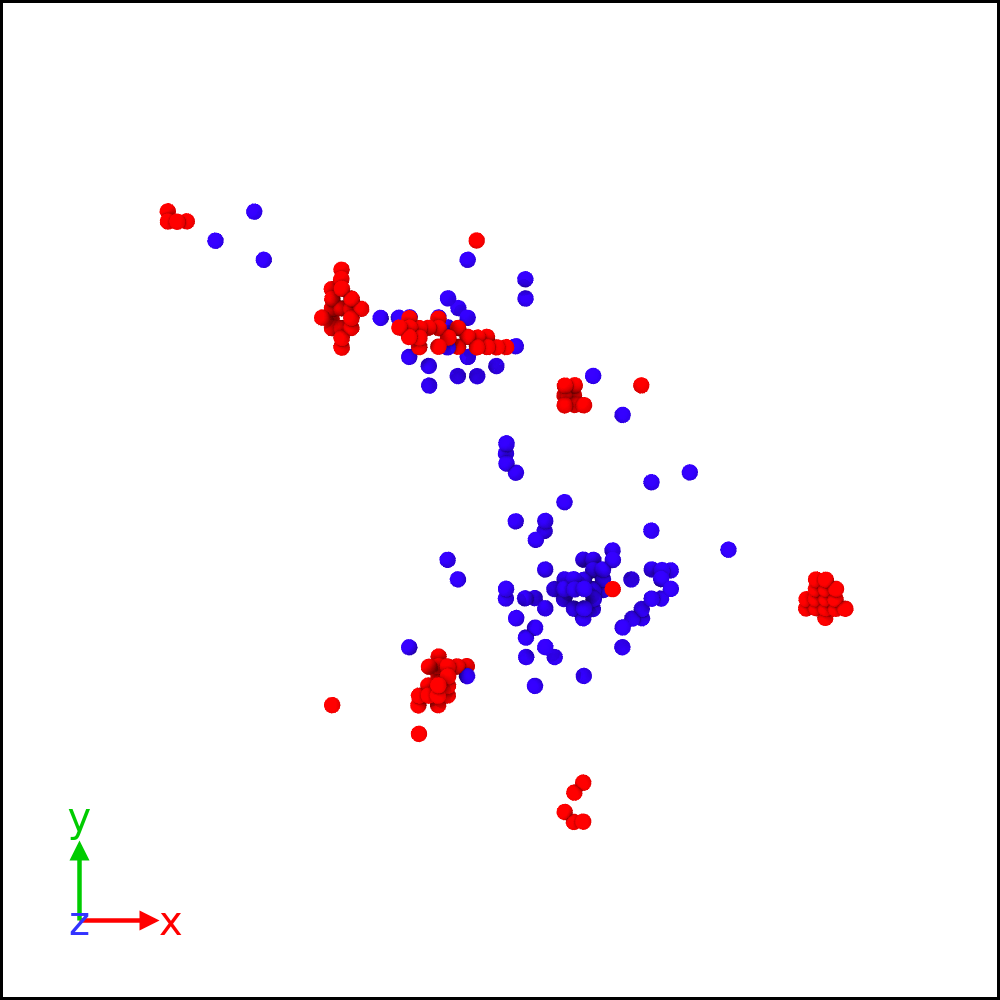}}
	\hspace{1em}
	\subfloat[]{\label{fig:nonttm_fp}\includegraphics[width=0.4\textwidth]{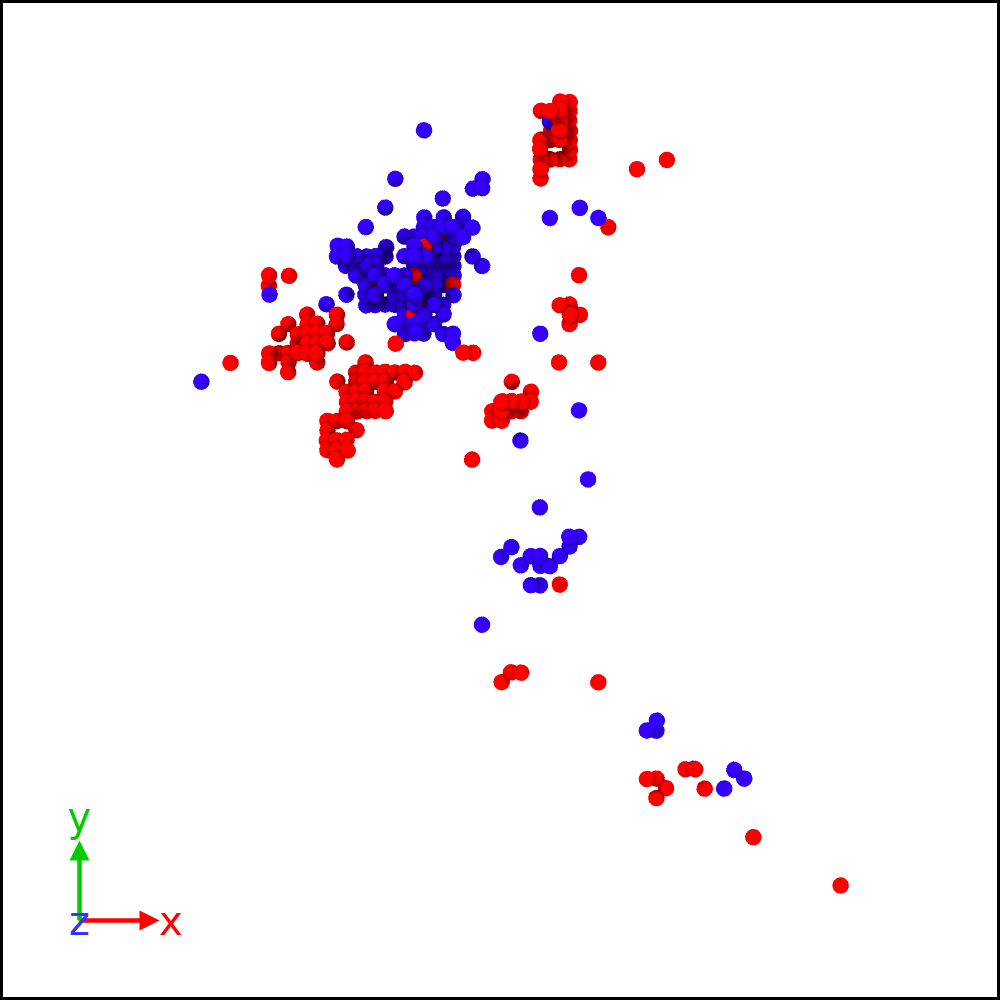}}
	\caption{Snapshot of point defects of a representative 50 KeV recoil simulation. Vacancies and interstitials are represented by blue and red color circles, respectively. (a) 2T-MD , (b) Classical MD}
	\label{fp_def}
\end{figure}

Next, we focus our attention to the evolution of the damage in the sample. The Voronoi cell method was used to identify the final damage production \cite{Nor:1998}. In this method, Voronoi polyhedra were centered on each atom position in the ideal fcc crystal and compared with atoms in the defective crystal. Polyhedra with no atoms were labeled as vacancies and polyhedra with 2 or more atoms were labeled as interstitials. The total number of Frenkel pairs (FPs) are 90 $\pm$ 10 and 120 $\pm$ 12 for ${\ell}$2T-MD and classical MD cascade simulations, respectively. The number of FPs and the standard error of the mean is calculated from 7 simulations for each simulation condition. We see that even though the time evolution of the kinetic energy in the ${\ell}$2T-MD model has very small differences with regular MD, the inclusion of electronic effects reduces the generation of FPs by 25\% with respect to MD indicating that electronic effects should be taken into account when studying irradiation damage. The low number of FPs in ${\ell}$2T-MD simulation is expected due to energy transfer to the electronic subsystem as shown in the kinetic energy plot (Fig. \ref{mean_temp} (b)). Figure \ref{fp_def} shows the final point defects of a representative classical MD simulation and the ${\ell}$2T-MD simulation having same recoil direction. The features of cascade damage are well captured in the simulation. It is visible from the snapshot that the interstitials are more aggregated due to high mobility and diffusivity compared to vacancies.

\section{Laser ablation}
Laser simulations were performed by giving a Neumann boundary conditions to the simulation cell. This boundary condition specifies the rate of change of the electronic temperature by giving a heat flow that follows the following expression \cite{iva:2003}
\begin{equation} \label{eq:LaserIntesity}
	q_{i} = i_0 \exp\left(\frac{-(x_i-x_0)}{l_p}\right)\exp\left(\frac{-(t-t_0)^2}{2\sigma^2} \right).
\end{equation}
Here, $i_0 = i(1-r)$ is the maximum absorbed intensity [temperature$\cdot$time$^{-1}$], $(1-r)$ reflects the amount of energy absorbed, and $r$ is the reflection parameter, $x_i$ and $x_0$ are the first components of the position vector of the $i^{\text{th}}$-atom and the free-surface end, where the laser pulse is applied, respectively. $l_p$ is the penetration length, $t_0$ is the time at which the pulse is applied and $\sigma$ is the standard deviation of the Gaussian profile. 

%

In our simulations, we selected  $i_0 = 10^5$ K$\cdot$ps$^{-1}$ (which gives a laser fluency of $F \approx 57$ J$\cdot$m$^{-2}$), the half-width duration and the duration of the pulse were  $\sigma = 1$ ps, and $\tau = 10.0$ ps, respectively. The penetration length was selected to be $l_p = 2$ nm. We notice that these parameters are arbitrarily selected for purposes of the example, and they might not be the best parameters to describe a particular laser wavelength. The simulation cell was $40 a_0\times40 a_0\times150 a_0 $ containing $\sim$960,000 atoms and it was carried out with a constant time step of $\Delta t = 5 \times 10^{-4}$ ps while the electronic time step was 10 times smaller. The material's constants and potential were specified in Section \ref{Section4}.

\begin{figure}[]
	\centering\captionsetup[subfloat]{labelfont=bf}
	\includegraphics[width=0.7\textwidth]{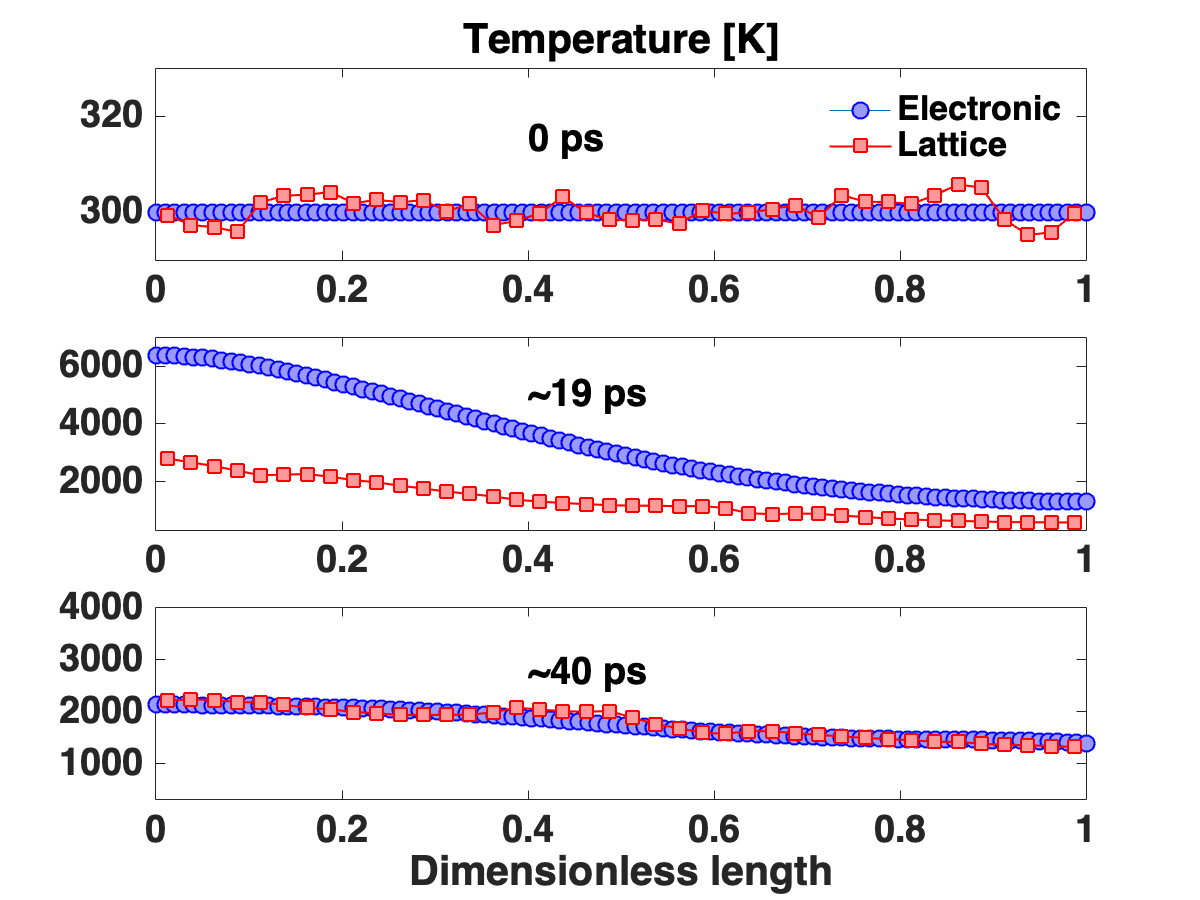}
	\caption{ Time evolution of the electronic and lattice temperature in the sample as a function of the position in the sample during the laser ablation simulation.}
	\label{Fig:LaserTemp}
\end{figure}

We first describe the time evolution of the lattice and electronic temperature across the sample. We see in Fig. \ref{Fig:LaserTemp}(a) that the lattice and electronic temperature are in equilibrium at $t=0$ ps. Some fluctuations appear in the lattice temperature due to the statistical nature of it. However, as the laser pulse is applied to the sample, the electronic temperature rises very rapidly in comparison with the lattice temperature, generating a state of thermal non-equilibrium. This can be seen  in Fig. \ref{Fig:LaserTemp}(b) where the electronic temperature is much higher than the lattice temperature. At $t=40$ ps, both electronic and lattice temperature reach an equilibrium again. However, we see that the left hand side of the bar is at a higher temperature than the right side. This is due to the fact that the bar has split into two parts, as described below.

\begin{figure}[!ht]
	\centering
	\subfloat[]{\label{fig:temp2ps}\includegraphics[width=0.33\textwidth]{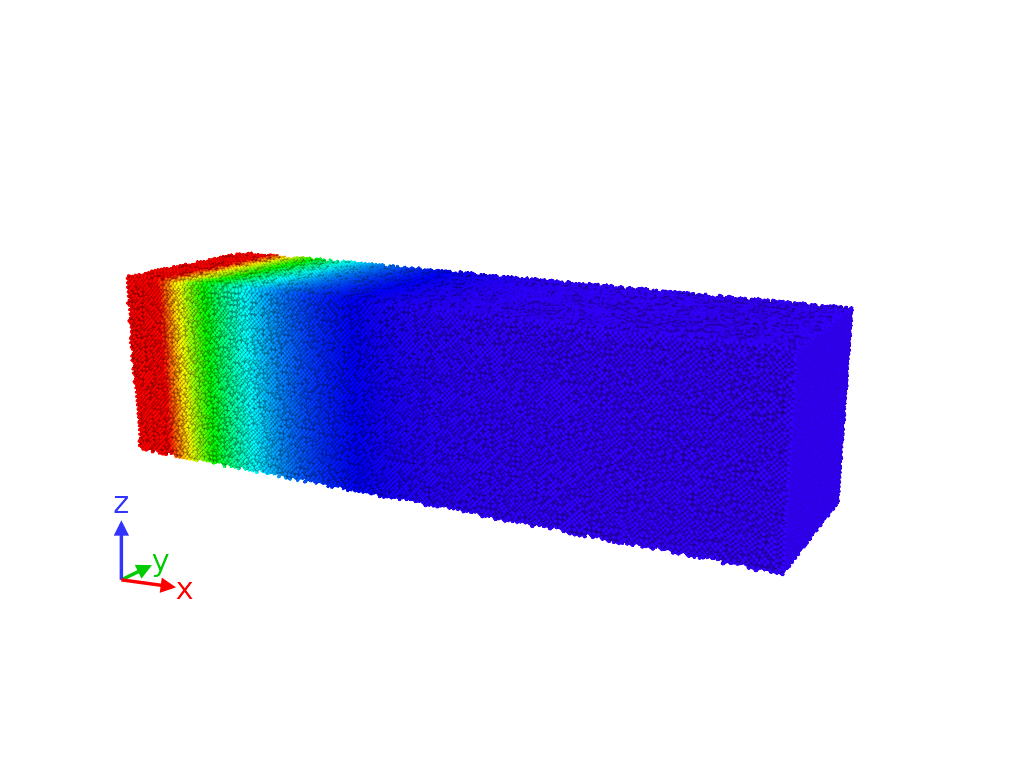}}
	\subfloat[]{\label{fig:temp2ps}\includegraphics[width=0.33\textwidth]{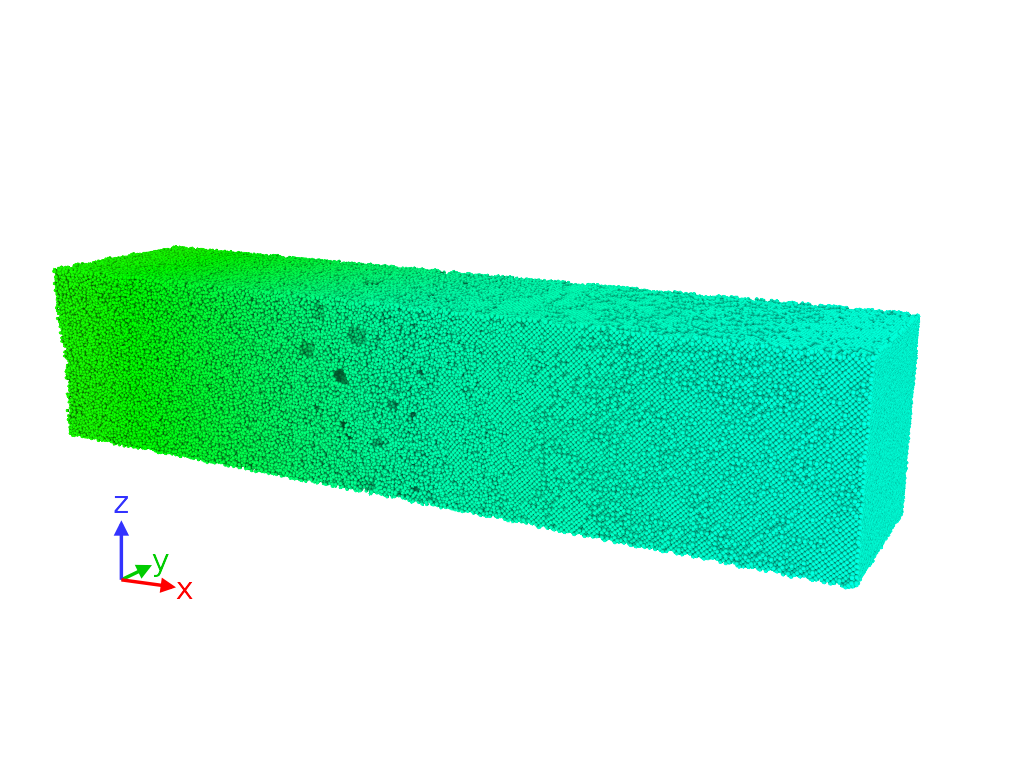}}
	\subfloat[]{\label{fig:temp2ps}\includegraphics[width=0.33\textwidth]{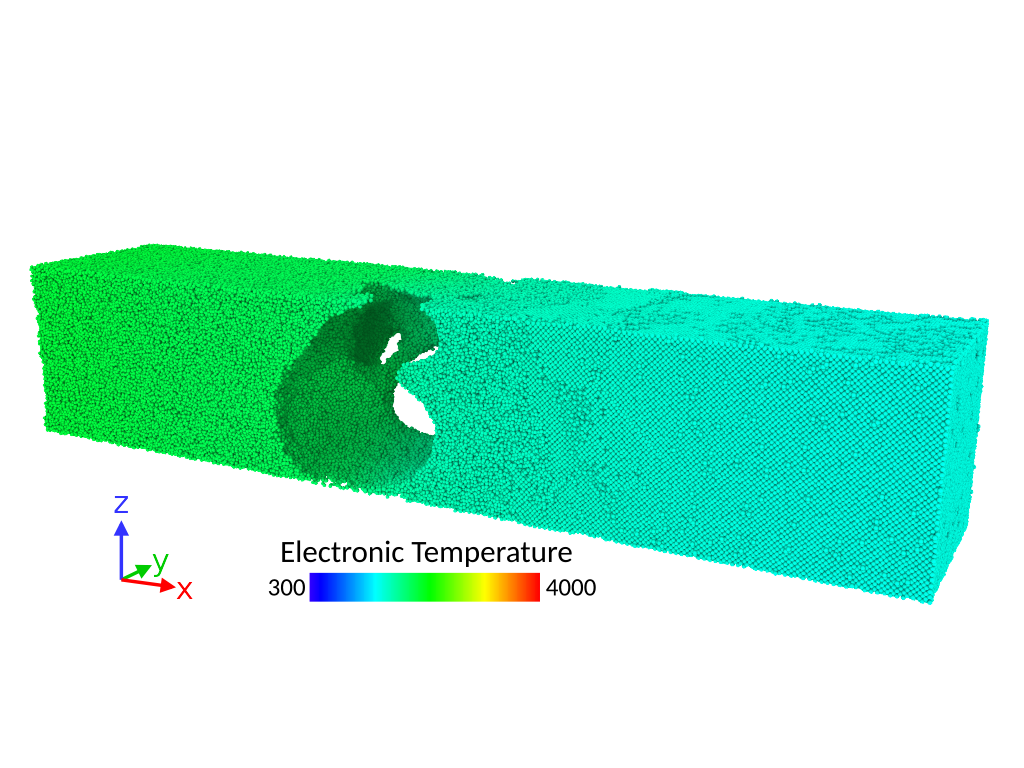}} 
	\caption{Temperature and atomic fraction of atoms for the laser pulse simulation. (a-c) Time evolution of the electronic temperature in the sample. (d-f) Time evolution of the atoms colored according to their local atomic structure. Amorphous (white), fcc (green), hcp (red) and bcc (blue).  }
	\label{Fig:LaserAblation}
\end{figure}

\begin{figure}[!ht]
	\centering
	\includegraphics[width=0.45\textwidth]{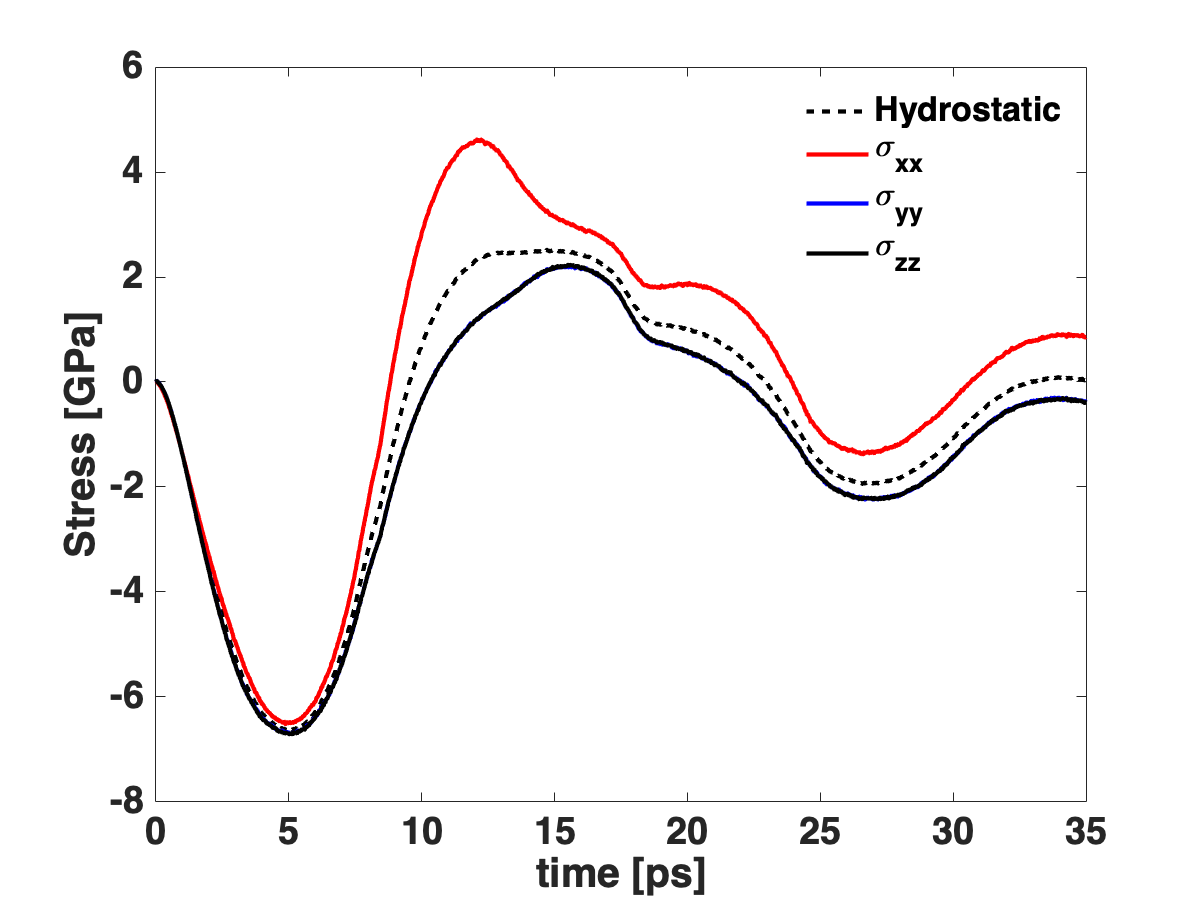}
	\caption{Time evolution of the hydrostatic pressure $\sigma_H$ in the sample, and the normal stresses.}
	\label{Fig:LaserPressure}
\end{figure}

Next, we describe the configurational changes suffered by the specimen during the process of laser ablation. In order to do so, we show the spatial configuration of the atoms for different times in Fig. \ref{Fig:LaserAblation}. Figs. \ref{Fig:LaserAblation}(a)-(c) show the spatial evolution of the temperature for $t = 0.1$, 19 and 40 ps, respectively. From the figure, we can see a large temperature gradient at time  $t = 0.1$. However, as time goes by, the temperature gradient decreases until stabilizes for $t= 40$ ps. Remarkably, we see the nucleation of small voids in Fig. \ref{Fig:LaserAblation}(b), which tend to open up to lead to a complete separation of the specimen as shown in Fig. \ref{Fig:LaserAblation}(c). In order to have a better understanding of this process, we investigated the evolution of the virial stress, in particular the hydrostatic component of it given by $\sigma_H = \frac{1}{3} (\sigma_{xx}+\sigma_{yy}+\sigma_{zz})$. This is shown in Fig. \ref{Fig:LaserPressure} where the hydrostatic and cartesian components of the stress tensor are shown.

Fig. \ref{Fig:LaserPressure} indicates that, initially, the laser pulse generates an overall compressive stress in the sample. This was evident in our simulations since the overall length of the sample significantly decreased due to the large compressive stress. The compressive pulse has its peak of -7 GPa at $t \approx 5$ ps. However, shortly after the peak compressive stress was reached, the stress reduced and changed its sign at around $t = 7$ ps. The maximum tensile value of 5 GPa was reached at around $t \approx 12$ ps. It is about this time where the voids start to nucleate due to the large hydrostatic stress in the sample. After this time, the stress fluctuates with values that are bounded by the maximum compressive and tensile values. This opened up large voids that eventually coalesced and split the sample in two parts.

The evolution of the local structure of the atoms was also monitored using the common neighbor analysis algorithm \cite{CNA:Stukowski}. At $t = 0$ ps, the sample contained mainly fcc atoms (99.9\%) and few atoms in other types due to surface effects (0.1\%). As the laser pulse gives energy to the sample, the fraction of atoms in amorphous phase increases almost linearly until $\approx 62\%$ of the atoms becomes amorphous at $t = 12$ ps, which coincides with the maximum peak stress. Thereafter, the portion of the atoms in amorphous phase remains the same, and mainly located in the left hand side of the images shown in Fig. \ref{Fig:LaserAblation}, where the laser pulse was applied.

\section{Conclusions}
We have developed, implemented, and validated a new version, called  (${\ell}$2T-MD), of the two temperature model coupled with MD simulations into the LAMMPS code. The methodology uses a master equation to compute effective rates of energy exchange in the electronic subsystem and it is coupled to the phonon vibrations of the atoms through classical MD. We have demonstrated the new methodology through multiple examples of technological relevance. 

In the cascade simulations, we found that while the mean average electronic temperature remains relatively small, the maximum temperature in the system can reach exceedingly large values. For instance, in the 50 KeV simulation, the mean electronic temperature increases up to $\sim$450 K, while the maximum reaches about $\sim$8,000 K. We also showed that the temperature gradients can be very large, of the order of $\sim$250 K$\cdot$nm$^{-1}$, as demonstrated in the electronic temperature across a section of the simulation. Such a large gradients, illustrates the ability of the master equation and the current approach to transport energy at the atomic scale including situations where the system is far from the thermodynamic equilibrium while retaining the smallest discretization possible in classical atomistic systems. Interestingly, the introduction of electrons in the irradiation damage simulations leads to a significant reduction of the numbers of Frenkel pairs with respect to MD, illustrating the need for including electronic effects. We also illustrated a laser ablation simulation, where a large amount of energy was introduced in the sample, generating melting of the sample, nanovoid nucleation, growth and coalescence of voids, leading to separation of the material. 

The current approach provides a mesh free alternative for coupling electrons with phonons in MD softwares. This is important because the implementation can be seamlessly implemented in multiple MD codes. We have done so using the LAMMPS code, providing capabilities for simulating atomic systems with massively parallel clusters. The implementation can be downloaded from its web site \cite{l2T-MD:website}. Future directions include the coupling of phonons with magnons to simulate ferroelectric materials.

\section{Acknowledgments}
We gratefully acknowledge the support from the Natural Sciences and Engineering Research Council of Canada (NSERC) through the Discovery Grant under Award Application Number RGPIN-2016-06114 and the support of Compute Canada through the Westgrid consortium. 



\end{document}